\documentclass[twocolumn,english,prb,showpacs]{revtex4}
\usepackage[T1]{fontenc}
\usepackage[latin1]{inputenc}
\usepackage{amsmath}
\usepackage{graphicx}
\usepackage{amssymb}

\makeatletter

\newcommand{\noun}[1]{\textsc{#1}}
\providecommand{\tabularnewline}{\\}

\usepackage{babel}
\makeatother
\begin{document}

\title{Efficient implementation of the \emph{GW} approximation within the
all-electron FLAPW method}

\author{Christoph Friedrich}

\email{c.friedrich@fz-juelich.de}

\author{Stefan Blügel}

\affiliation{Institut für Festkörperforschung and Institute for Advanced Simulation,
Forschungszentrum Jülich, 52425 Jülich, Germany}

\author{Arno Schindlmayr}

\affiliation{Department Physik, Universität Paderborn, 33095 Paderborn, Germany}

\begin{abstract}
We present an implementation of the $GW$ approximation for the electronic
self-energy within the full-potential linearized augmented-plane-wave
(FLAPW) method. The algorithm uses an all-electron mixed product basis
for the representation of response matrices and related quantities.
This basis is derived from the FLAPW basis and is exact for wave-function
products. The correlation part of the self-energy is calculated on
the imaginary frequency axis with a subsequent analytic continuation
to the real axis. As an alternative we can perform the frequency convolution
of the Green function $G$ and the dynamically screened Coulomb interaction
$W$ explicitly by a contour integration. The singularity of the bare
and screened interaction potentials gives rise to a numerically important
self-energy contribution, which we treat analytically to achieve good
convergence with respect to the $\mathbf{k}$-point sampling. As numerical
realizations of the $GW$ approximation typically suffer from the
high computational expense required for the evaluation of the nonlocal
and frequency-dependent self-energy, we demonstrate how the algorithm
can be made very efficient by exploiting spatial and time-reversal
symmetry as well as by applying an optimization of the mixed product
basis that retains only the numerically important contributions of
the electron-electron interaction. This optimization step reduces
the basis size without compromising the accuracy and accelerates the
code considerably. Furthermore, we demonstrate that one can employ
an extrapolar approximation for high-lying states to reduce the number
of empty states that must be taken into account explicitly in the
construction of the polarization function and the self-energy. We
show convergence tests, CPU timings, and results for prototype semiconductors
and insulators as well as ferromagnetic nickel.
\end{abstract}

\pacs{71.15.Qe, 71.20.Mq, 71.45.Gm}

\maketitle

\section{Introduction}

Many-body perturbation theory with the $GW$ approximation for the
electronic self-energy offers a well-established approach for the
computation of excited electronic states from first principles.\cite{Aulbur2000}
In principle, the electronic self-energy incorporates all many-body
exchange and correlation effects beyond the Hartree theory. The $GW$
approximation contains the electronic exchange exactly while the screening
is treated at the level of the random-phase approximation,\cite{Adler1962}
where noninteracting electron-hole ring diagrams are summed to all
orders. This makes the $GW$ approximation particularly suited for
weakly to moderately correlated systems.

After its theoretical foundation by Hedin \cite{Hedin1965} in 1965
it was not before the middle of the 1980s that the computational treatment
of real materials became feasible. In spite of approximations in the
numerical treatment that were necessary due to the lack of computer
power, the first results were very promising. In these works, \cite{Godby1986,Godby1987,Hybertsen1985,Hybertsen1986}
it could be shown that the theoretical band gap fell within a margin
of 0.1~eV from the experimental value for covalently bonded semiconductors.
After these pioneering studies, the $GW$ approximation has evolved
into the method of choice for calculating electronic excitations in
solid-state systems.

So far, most codes still rely on the pseudopotential approximation,
which restricts the range of materials that can be examined. Transition-metal
compounds and oxides, in particular, cannot be treated efficiently
in this approach. Two early all-electron calculations using the $GW$
approximation were carried out by Hamada \emph{et al.}\cite{Hamada1990}
for Si and by Aryasetiawan \cite{Aryasetiawan1992} for Ni, both within
the linearized augmented-plane-wave (LAPW) method. \emph{}However,
only very recently were further all-electron implementations reported,
based on the full-potential LAPW (FLAPW) (Refs.~\onlinecite{Ku2002}
and \onlinecite{Usuda2002}), the linearized muffin-tin orbital (LMTO)
(Refs.~\onlinecite{Kotani2002}--\onlinecite{Faleev2004}), the projector-augmented-wave
(PAW) (Refs.~\onlinecite{Arnaud2000}--\onlinecite{Shishkin2006}),
and the Korringa-Kohn-Rostoker method (Ref.~\onlinecite{Ernst2005})
together with applications to a larger variety of systems. 

While the calculation of $GW$ band structures for small systems has
become routine, the scientific community is increasingly interested
in larger and more complex systems, such as multicomponent materials,
artificial heterostructures, defects, interfaces, surfaces, clusters,
and nanowires. In codes using periodic boundary conditions, such systems
must be treated in supercell geometries often exceeding 100 atoms.
The main obstacle in applying the $GW$ approximation in supercell
calculations is the considerable demand of computation time and memory.
This is especially true for all-electron methods, where the rapid
oscillations close to the atomic nuclei make the usage of the fast
Fourier transformation impossible. Therefore, all-electron $GW$ calculations
for large systems have so far been prohibitive. In this paper, we
describe numerical algorithms and approximations that make all-electron
$GW$ implementations efficient and bring large supercell calculations
into reach.

The nonlocality of the self-energy operator is the main reason for
the large computational effort needed in $GW$ calculations. It leads
to convolutions in reciprocal space, i.e., summations over the Brillouin
zone (BZ), which is sampled by a finite set of $\mathbf{k}$ points.
Furthermore, the bare electron-electron interaction diverges at the
center of the BZ, giving rise to a corresponding divergence and an
anisotropy in the dynamically screened interaction at $\mathbf{k=0}$.
A thorough treatment of the $\Gamma$ point is hence crucial for an
accurate and efficient BZ summation. Previous all-electron implementations
\cite{Ku2002,Puschnig2002} of many-body perturbation theory have
often bypassed this problem by reverting to a plane-wave basis for
the Coulomb potential and related propagators such as the dielectric
function but such a projection leads to a loss of accuracy because
it cannot resolve the rapid oscillations of the orbitals close to
the atomic nuclei without a prohibitively large basis set. As a consequence,
physical effects such as core polarization are inadequately described.
In an alternative approach, the so-called offset-$\Gamma$ method,
an auxiliary $\mathbf{k}$-point mesh that is shifted from the origin
by a small but finite amount is employed.\cite{Kotani2002,Lebegue2003}
In this way, the singularity is avoided but the use of additional
meshes increases the numerical cost; even in the most favorable case,
for cubic symmetry, the number of $\mathbf{k}$ points must at least
be doubled. Furthermore, the convergence of BZ integrals involving
the Coulomb matrix, for example, for the $GW$ self-energy, may be
slow with respect to the $\mathbf{k}$-point sampling due to the approximate
treatment of the quantitatively important region near the zone center.
In this work, we take the $\Gamma$ point explicitly into account
and employ an analytical treatment of the Coulomb singularity. We
use a recently developed procedure\cite{Friedrich2009} to transform
the all-electron basis for the interaction potentials, the so-called
mixed product basis,\cite{Kotani2002} in such a way that the divergence
is restricted to a single matrix element, which allows a treatment
similar to a pure plane-wave basis set. 

The numerical cost of the BZ convolutions can be reduced considerably
by employing spatial and time-reversal symmetries. Not only can the
$\mathbf{k}$ dependence of response quantities be confined to the
irreducible wedge of the BZ but the also convolutions over the reciprocal
space can be restricted to an extended irreducible zone without loss
of accuracy. Furthermore, in the presence of inversion symmetry, the
all-electron mixed product basis can be transformed in such a way
that response matrices become real instead of complex, which again
reduces the numerical cost in terms of computation time and memory
demand. 

We further demonstrate that we can afford to truncate the number of
basis functions considerably in the calculation of the correlation
part of the self-energy. This is achieved by a basis transformation
that diagonalizes the Coulomb matrix. Eigenvectors with small eigenvalues
then correspond to unimportant scattering contributions, which can
be neglected in a systematic way. This leads to an optimization of
the basis set and thus to a speed up of the computation. 

The paper is organized as follows. In Sec.~\ref{sec:GWapprox}, we
give a brief introduction to the $GW$ approximation. In Sec.~\ref{sec:Implementation},
we describe our all-electron implementation in detail: the mixed product
basis and its optimization, the $\mathbf{k}$-point set, the treatment
of the $\Gamma$-point divergence, and the usage of spatial and time-reversal
symmetry. Section \ref{sec:Test} reports convergence tests for Si
and $\mathrm{SrTiO}_{3}$ as well as fundamental $GW$ band gaps for
a variety of semiconductors and insulators. In addition, results for
the localized $3d$ states of GaAs and ferromagnetic Ni as an example
of a $3d$ transition metal are discussed. In order to illustrate
the efficiency of the code, we also show CPU timings for diamond in
supercell geometries containing up to 128 atoms. Finally we summarize
our conclusions in Sec.~\ref{sec:Conclusions}.

\section{$\textrm{\boldmath$GW$}$ approximation\label{sec:GWapprox}}

Angle-resolved photoelectron spectroscopy is the prime experimental
technique for the measurement of the electronic band structure of
crystalline materials. The excitations measured in photoelectron spectroscopy
involve electron ejection or injection and thus imply a change in
the particle number by one. The corresponding many-body excitation
energies $E_{n\mathbf{q}}^{\sigma}$, where $\mathbf{q}$ is the wave
vector, $n$ the band index, and $\sigma$ the electron spin, define
the pole structure of the one-particle Green function,\begin{equation}
G^{\sigma}(\mathbf{r},\mathbf{r}';\omega)=\sum_{\mathbf{q}}^{\mathrm{BZ}}\sum_{n}^{\mathrm{all}}\frac{\psi_{n\mathbf{q}}^{\sigma}(\mathbf{r})\psi_{n\mathbf{q}}^{\sigma^{{\scriptstyle *}}}(\mathbf{r}')}{\omega-E_{n\mathbf{q}}^{\sigma}+E_{0}+i\eta\,\textrm{sgn}(E_{n\mathbf{q}}^{\sigma}-E_{0})}\label{eq:GreenFunction}\end{equation}
with the quasiparticle wave functions $\psi_{n\mathbf{q}}^{\sigma}(\mathbf{r})$
and the many-body ground-state energy $E_{0}$. Here and in the following
the number $\eta$ is infinitesimal, real, and positive, and by a
sum over Bloch vectors $\mathbf{q}$, we always mean an integration
over the BZ multiplied by the density of $\mathbf{q}$ points $V/(8\pi^{3})$
with the crystal volume $V$. Hartree atomic units are used throughout
except where noted otherwise. 

The quasiparticle wave functions $\psi_{n\mathbf{q}}^{\sigma}(\mathbf{r})$
and energies $E_{n\mathbf{q}}^{\sigma}$ obey a set of one-particle
differential equations,\begin{eqnarray}
\hat{h}_{0}\psi_{n\mathbf{q}}^{\sigma}(\mathbf{r})+\int\left[\Sigma_{\mathrm{xc}}^{\sigma}(\mathbf{r},\mathbf{r}';E_{n\mathbf{q}})\right.\nonumber \\
\left.-v_{\mathrm{xc}}^{\sigma}(\mathbf{r}')\delta(\mathbf{r}-\mathbf{r}')\right]\psi_{n\mathbf{q}}^{\sigma}(\mathbf{r}')\, d^{3}r' & = & E_{n\mathbf{q}}^{\sigma}\psi_{n\mathbf{q}}^{\sigma}(\mathbf{r})\,,\label{eq:quasiparteq}\end{eqnarray}
the so-called quasiparticle equations, where $\hat{h}_{0}$ is the
Kohn-Sham (KS) Hamiltonian\begin{equation}
\hat{h}_{0}=-\frac{1}{2}\nabla^{2}+v_{\mathrm{ext}}(\mathbf{r})+v_{\mathrm{H}}(\mathbf{r})+v_{\mathrm{xc}}^{\sigma}(\mathbf{r})\end{equation}
with the external, Hartree, and exchange-correlation potentials, respectively,
\cite{Kohn1965} and $\Sigma_{\mathrm{xc}}^{\sigma}(\mathbf{r},\mathbf{r}';\omega)$
is the nonlocal, non-Hermitian, and energy-dependent exchange-correlation
self-energy.

In practical implementations one usually treats the integral operator
on the left-hand side of Eq.~(\ref{eq:quasiparteq}) as a small perturbation.
In first order, the quasiparticle energies are then given by the nonlinear
equation,\begin{equation}
E_{n\mathbf{q}}^{\sigma}=\epsilon_{n\mathbf{q}}^{\sigma}+\left\langle \varphi_{n\mathbf{q}}^{\sigma}\right|\Sigma_{\mathrm{xc}}^{\sigma}(E_{n\mathbf{q}}^{\sigma})-v_{\mathrm{xc}}^{\sigma}\left|\varphi_{n\mathbf{q}}^{\sigma}\right\rangle \label{eq:quasipart_corr}\end{equation}
with the KS wave functions $\varphi_{n\mathbf{q}}^{\sigma}$ and energies
$\epsilon_{n\mathbf{q}}^{\sigma}$. This is equivalent to neglecting
off-diagonal elements of $\Sigma_{\mathrm{xc}}^{\sigma}(E_{n\mathbf{q}}^{\sigma})-v_{\mathrm{xc}}^{\sigma}$
in the basis of the KS wave functions. For the self-energy, we use
the $GW$ approximation,\begin{equation}
\Sigma_{\mathrm{xc}}^{\sigma}(\mathbf{r},\mathbf{r}';\omega)=\frac{i}{2\pi}\int G_{0}^{\sigma}(\mathbf{r},\mathbf{r}';\omega+\omega')W(\mathbf{r},\mathbf{r}';\omega')e^{i\eta\omega'}d\omega'\,,\label{eq:selfeneGW}\end{equation}
which constitutes the first-order term of the self-energy expansion
in powers of the dynamically screened Coulomb interaction $W$. Here\begin{equation}
G_{0}^{\sigma}(\mathbf{r},\mathbf{r}';\omega)=\sum_{\mathbf{q}}^{\mathrm{BZ}}\sum_{n}^{\mathrm{all}}\frac{\varphi_{n\mathbf{q}}^{\sigma}(\mathbf{r})\varphi_{n\mathbf{q}}^{\sigma^{{\scriptstyle *}}}(\mathbf{r}')}{\omega-\epsilon_{n\mathbf{q}}^{\sigma}+i\eta\,\textrm{sgn}(\epsilon_{n\mathbf{q}}^{\sigma})}\,.\label{eq:Green0}\end{equation}
 is the time-ordered KS Green function, which is obtained from Eq.~(\ref{eq:GreenFunction})
by replacing $\psi_{n\mathbf{q}}^{\sigma}(\mathbf{r})$ with the KS
wave functions $\varphi_{n\mathbf{q}}^{\sigma}(\mathbf{r})$ and $E_{n\mathbf{q}}^{\sigma}-E_{0}$
with the KS energies $\epsilon_{n\mathbf{q}}^{\sigma}$ measured from
the Fermi energy.

The dynamically screened interaction obeys a Dyson-type integral equation,\begin{eqnarray}
\lefteqn{W(\mathbf{r},\mathbf{r}';\omega)=v(\mathbf{r},\mathbf{r}')}\label{eq:W_Dyson}\\
 &  & +\int v(\mathbf{r},\mathbf{r}'')P(\mathbf{r}'',\mathbf{r}''';\omega)W(\mathbf{r}''',\mathbf{r}';\omega)\, d^{3}r''\, d^{3}r'''\nonumber \end{eqnarray}
with the bare Coulomb interaction $v(\mathbf{r},\mathbf{r}')=1/\left|\mathbf{r}-\mathbf{r}'\right|$
and the polarization function $P(\mathbf{r},\mathbf{r}';\omega)$,
for which we employ the random-phase approximation, \cite{Adler1962}\begin{eqnarray}
\lefteqn{P(\mathbf{r},\mathbf{r}';\omega)}\label{eq:polar_r}\\
 & = & -\frac{i}{2\pi}\sum_{\sigma}\int_{-\infty}^{\infty}G_{0}^{\sigma}(\mathbf{r},\mathbf{r}';\omega')G_{0}^{\sigma}(\mathbf{r}',\mathbf{r};\omega'-\omega)e^{i\eta\omega'}\, d\omega'\nonumber \\
 & = & \sum_{\sigma}\sum_{\mathbf{q},\mathbf{k}}^{\mathrm{BZ}}\sum_{n}^{\mathrm{occ}}\sum_{n'}^{\mathrm{unocc}}\varphi_{n\mathbf{q}}^{\sigma^{{\scriptstyle *}}}(\mathbf{r})\varphi_{n'\mathbf{k}}^{\sigma}(\mathbf{r})\varphi_{n\mathbf{q}}^{\sigma}(\mathbf{r}')\varphi_{n'\mathbf{k}}^{\sigma^{{\scriptstyle *}}}(\mathbf{r}')\nonumber \\
 &  & \times\left(\frac{1}{\omega+\epsilon_{n\mathbf{q}}^{\sigma}-\epsilon_{n'\mathbf{k}}^{\sigma}+i\eta}-\frac{1}{\omega-\epsilon_{n\mathbf{q}}^{\sigma}+\epsilon_{n'\mathbf{k}}^{\sigma}-i\eta}\right)\,.\nonumber \end{eqnarray}
This approximation corresponds to time-dependent Hartree theory and
thus neglects exchange-correlation (e.g., excitonic) effects in the
dynamical screening. With the dielectric function,\begin{equation}
\varepsilon(\mathbf{r},\mathbf{r}';\omega)=\delta(\mathbf{r}-\mathbf{r}')-\int v(\mathbf{r},\mathbf{r}'')P(\mathbf{r}'',\mathbf{r}';\omega)\, d^{3}r''\label{eq:dielec}\end{equation}
we can write the screened interaction in the closed form,\begin{equation}
W(\mathbf{r},\mathbf{r}';\omega)=\int\varepsilon^{-1}(\mathbf{r},\mathbf{r}'';\omega)v(\mathbf{r}'',\mathbf{r}')\, d^{3}r''\,.\label{eq:screen_eps}\end{equation}
Many-body screening effects obviously enter with the second term of
Eq.~(\ref{eq:dielec}) into the formalism. In fact, if we write the
screened interaction as a sum of the bare interaction and a remainder,\begin{equation}
W(\mathbf{r},\mathbf{r}';\omega)=v(\mathbf{r},\mathbf{r}')+W^{\mathrm{c}}(\mathbf{r},\mathbf{r}';\omega)\,,\end{equation}
then the self-energy {[}Eq.~(\ref{eq:selfeneGW}){]} decomposes into
the terms \begin{equation}
\Sigma_{\mathrm{x}}^{\sigma}(\mathbf{r},\mathbf{r}')=\frac{i}{2\pi}\int G_{0}^{\sigma}(\mathbf{r},\mathbf{r}';\omega+\omega')v(\mathbf{r},\mathbf{r}')e^{i\eta\omega'}d\omega'\label{eq:selfeneHF_r}\end{equation}
and\begin{equation}
\Sigma_{\mathrm{c}}^{\sigma}(\mathbf{r},\mathbf{r}';\omega)=\frac{i}{2\pi}\int G_{0}^{\sigma}(\mathbf{r},\mathbf{r}';\omega+\omega')W^{\mathrm{c}}(\mathbf{r},\mathbf{r}';\omega')\, d\omega'\,,\label{eq:selfenecorr_r}\end{equation}
 which are identified as the exchange and the correlation contributions
to the electronic self-energy, respectively. We note that the exponential
factor allows to close the integration path over the upper complex
half plane in Eq.~(\ref{eq:selfeneHF_r}). As $W^{\mathrm{c}}(\mathbf{r},\mathbf{r}';\omega)$
falls off quickly enough with increasing frequencies, we may take
the limit $\eta\rightarrow0$ before integrating in Eq.~(\ref{eq:selfenecorr_r}). 

In the next section, we discuss several aspects of the implementation
that are important for the computational efficiency. The numerical
procedure is based on an auxiliary all-electron basis set, the mixed
product basis, in which the previous integral equations become matrix
equations that can be implemented easily. We have already explained
this basis set in detail in a previous publication \cite{Friedrich2009}
and only sketch the basic ideas here.

\section{Implementation\label{sec:Implementation}}

\subsection{FLAPW method}

In the FLAPW method,\cite{Andersen1975} space is partitioned into
nonoverlapping atom-centered muffin-tin (MT) spheres and the interstitial
region (IR). The core-electron wave functions, which are (predominantly)
confined to the MT spheres, are directly obtained from a solution
of the fully relativistic Dirac equation. The valence-electron wave
functions with spin $\sigma$ are expanded in interstitial plane waves
(IPWs) in the IR and numerical functions $u_{almp}^{\sigma}(\mathbf{r})=u_{alp}^{\sigma}(r)Y_{lm}(\mathbf{e_{r}})$
inside the MT sphere of atom $a$, where $\mathbf{r}$ is measured
from the MT center located at $\mathbf{R}_{a}$. These numerical functions
comprise solutions of the KS equation for the spherically averaged
effective potential for $p=0$ and their first energy derivatives
$u_{alm1}^{\sigma}(\mathbf{r})=\partial u_{alm0}^{\sigma}(\mathbf{r})/\partial\epsilon_{al}^{\sigma}$
for $p=1$ evaluated at suitably chosen energy parameters $\epsilon_{al}^{\sigma}$,
and $Y_{lm}(\mathbf{e_{r}})$ denote the spherical harmonics. The
notation $\mathbf{e}_{\mathbf{r}}=\mathbf{r}/r$ with $r=|\mathbf{r}|$
indicates the unit vector in the direction of $\mathbf{r}$. In a
given unit cell, the KS wave function at a wave vector $\mathbf{k}$
with band index $n$ and spin $\sigma$ is then given by\begin{equation}
\varphi_{n\mathbf{k}}^{\sigma}(\mathbf{r})=\left\{ \begin{array}{l}
{\displaystyle \frac{1}{\sqrt{N}}\sum_{l=0}^{l_{\mathrm{max}}}\sum_{m=-l}^{l}\sum_{p=0}^{1}}A_{almp}^{n\mathbf{k}\sigma}u_{almp}^{\sigma}(\mathbf{r}-\mathbf{R}_{a})\\
\quad\quad\quad\quad\quad\quad\quad\quad\quad\quad\quad\quad\,\,\,\,\,\,\textrm{if }\mathbf{r}\in\mathrm{MT}(a)\\
{\displaystyle \frac{1}{\sqrt{V}}\sum_{|\mathbf{k+G}|\le G_{\mathrm{max}}}}c_{\mathbf{G}}^{n\mathbf{k}\sigma}e^{i(\mathbf{k}+\mathbf{G})\cdot\mathbf{r}}\textrm{ if }\mathbf{r}\in\mathrm{IR}\end{array}\right.\label{Eq:FLAPW-basis}\end{equation}
with the crystal volume $V$, the number of unit cells $N$, and cutoff
values $l_{\mathrm{max}}$ and $G_{\mathrm{max}}$. The coefficients
$A_{almp}^{n\mathbf{k}\sigma}$ are determined by the requirement
that the wave function is continuous in value and first radial derivative
at the MT sphere boundaries. If desired, additional local orbitals
for semicore states \cite{Singh1991} or higher energy derivatives
\cite{Friedrich2006} can be incorporated by allowing $p\ge2$. We
use the \noun{fleur} code\cite{Fleur} for the density-functional
theory (DFT) calculations.

\subsection{Mixed product basis (MPB)\label{sec:mixedbasis}}

If the integral equations of Sec.~\ref{sec:GWapprox} are rewritten
in an auxiliary basis set, they become matrix equations, which can
easily be treated in a computer code using standard linear-algebra
libraries. Equation (\ref{eq:polar_r}) already indicates that this
auxiliary basis set should accurately represent wave-function products.
This is generally true for all quantities that involve two spatial
coordinates and thus couple two incoming and outgoing electrons with
each other. 

The FLAPW method uses continuous basis functions that are defined
everywhere in space but have a different mathematical representation
in the MT spheres and the IR\@. For the expansion of wave-function
products, however, it is better to employ two separate sets of functions
that are defined only in one of the spatial regions and zero in the
other. In this way, linear dependences that occur only in one region
can easily be eliminated, which overall leads to a smaller and more
efficient basis. The resulting combined set of functions is called
the MPB.\cite{Kotani2002}

Inside the MT spheres, the MPB must accurately describe the products
$u_{almp}^{\sigma\,*}(\mathbf{r})u_{al'm'p'}^{\sigma}(\mathbf{r})$.
The angular parts $Y_{lm}^{*}(\mathbf{e_{r}})Y_{l'm'}(\mathbf{e_{r}})$
can be represented by linear combinations of spherical harmonics $Y_{LM}(\mathbf{e_{r}})$
with $|l-l'|\le L\le l+l'$ and $-L\le M\le L$, while the radial
parts define a set of product functions $U_{aLP}^{\sigma}(r)=u_{alp}^{\sigma}(r)u_{al'p'}^{\sigma}(r)$,
where the index $P$ counts all possible combinations of $l$, $l'$,
$p$, and $p'$. We emphasize that, in general, the latter lie outside
the vector space spanned by the original numerical basis functions
$u_{alp}^{\sigma}(r)$. Initially, the set $\{ U_{aLP}^{\sigma}(r)\}$
is neither normalized nor orthogonal and usually has a high degree
of (near) linear dependence. An effective procedure to remove these
(near) linear dependences is to diagonalize the overlap matrix and
to retain only those eigenvectors whose eigenvalues exceed a specified
threshold value. \cite{Aryasetiawan1994} In this way, the MT functions
become orthonormalized. By using both spin-up and spin-down products
in the construction of the overlap matrix, we make the resulting basis
spin-independent. In practice, the basis set is reduced further by
introducing a cutoff value $L_{\mathrm{max}}$ for the angular quantum
number. On the other hand, it must be supplemented with a constant
MT function for each atom in the unit cell, which is later needed
to represent the eigenfunction that corresponds to the divergent eigenvalue
of the Coulomb matrix in the limit $\mathbf{k}\rightarrow\mathbf{0}$.
From the resulting MT functions $M_{aLMP}(\mathbf{r})=M_{aLP}(r)Y_{LM}(\mathbf{e_{r}})$,
we formally construct Bloch functions. 

In the IR, we use a set of IPWs with a cutoff $G_{\mathrm{max}}^{\prime}\le2G_{\mathrm{max}}$
in reciprocal space, since the product of two IPWs yields another
IPW\@. Together with the MT functions, we thus obtain the MPB $\{ M_{I}^{\mathbf{k}}(\mathbf{r})\}=\{ M_{aLMP}^{\mathbf{k}}(\mathbf{r}),M_{\mathbf{G}}^{\mathbf{k}}(\mathbf{r})\}$
for the representation of wave-function products. Unlike the MT functions,
which have been explicitly orthonormalized, the IPWs are not orthogonal
to each other; the elements of their overlap matrix can be calculated
analytically and are given by\begin{equation}
\langle M_{\mathbf{G}}^{\mathbf{k}}|M_{\mathbf{G}'}^{\mathbf{k}'}\rangle=\delta_{\mathbf{kk}'}O_{\mathbf{GG}'}(\mathbf{k})=\delta_{\mathbf{kk}'}\Theta_{\mathbf{G}-\mathbf{G}'}\,,\end{equation}
where $\Theta_{\mathbf{G}}$ are the Fourier coefficients of the step
function, which equals 1 in the IR and 0 in the MT spheres. We also
define a second set, the biorthogonal set,\begin{equation}
\tilde{M}_{\mathbf{G}}^{\mathbf{k}}(\mathbf{r})=\sum_{\mathbf{G}'}M_{\mathbf{G}'}^{\mathbf{k}}(\mathbf{r})O_{\mathbf{G}'\mathbf{G}}^{-1}(\mathbf{k})\,.\label{eq:biorthog}\end{equation}
It fulfills the identities\begin{subequations}\label{eq:biorthog_ident}\begin{equation}
\langle\tilde{M}_{I}^{\mathbf{k}}|M_{J}^{\mathbf{k}}\rangle=\langle M_{I}^{\mathbf{k}}|\tilde{M}_{J}^{\mathbf{k}}\rangle=\delta_{IJ}\end{equation}
\begin{equation}
\sum_{I}|M_{I}^{\mathbf{k}}\rangle\langle\tilde{M}_{I}^{\mathbf{k}}|=\sum_{I}|\tilde{M}_{I}^{\mathbf{k}}\rangle\langle M_{I}^{\mathbf{k}}|=1\,,\end{equation}
\end{subequations}where the completeness relation is only valid in
the subspace spanned by the MPB, though. As the MT functions and the
IPWs are defined in different regions of space and the MT functions
are orthonormal, only the IPWs overlap in a nontrivial way. It should
be noted that the overlap matrix is $\mathbf{k}$-dependent because
the size of the MPB varies for different $\mathbf{k}$ vectors.

In general, the matrix representation of real operators in an arbitrary
complex basis $\{ f_{\mu}(\mathbf{r})\}$ is Hermitian. If the system
has inversion symmetry and the basis functions fulfill\begin{equation}
f_{\mu}(-\mathbf{r})=f_{\mu}^{*}(\mathbf{r})\,,\label{eq:condition_inv}\end{equation}
it is easy to show that the matrices become real symmetric. Of course,
this reduces the computational cost considerably, in terms of both
memory consumption and computation time. However, according to the
current definition only the IPWs fulfill Eq.~(\ref{eq:condition_inv})
while the MT functions do not. For a system with inversion symmetry,
we hence apply a unitary transformation of the MT functions such that
Eq.~(\ref{eq:condition_inv}) is satisfied.\cite{Betzinger} In the
following, it is understood that all quantities are represented in
this symmetrized basis if inversion symmetry is present.

\subsection{Formulation in the MPB\label{sub:Formulation-MPB}}

In this section, we reformulate the equations of Sec.~\ref{sec:GWapprox}
by projecting onto the MPB and exploiting the identities in Eq.~(\ref{eq:biorthog_ident}).
Because of the exponential factor in Eq.~(\ref{eq:selfeneHF_r}),
we can formally close the frequency integration contour with an infinite
half circle over the positive imaginary plane without changing the
value of the integral. This contour integral then equals the sum over
the residues of the poles of the Green function. The expectation value
of the exchange term with respect to a wave function $\varphi_{n\mathbf{q}}^{\sigma}(\mathbf{r})$
yields the well-known Hartree-Fock expression\begin{eqnarray}
\lefteqn{\langle\varphi_{n\mathbf{q}}^{\sigma}|\Sigma_{\mathrm{x}}^{\sigma}|\varphi_{n\mathbf{q}}^{\sigma}\rangle=-\sum_{\mathbf{k}}^{\mathrm{BZ}}\sum_{n'}^{\mathrm{occ.}}\sum_{I,J}v_{IJ}(\mathbf{k})}\label{eq:HFterm_MB}\\
 &  & \times\langle\varphi_{n'\mathbf{q+k}}^{\sigma}|\varphi_{n\mathbf{q}}^{\sigma}\tilde{M}_{I}^{\mathbf{k}}\rangle\langle\tilde{M}_{J}^{\mathbf{k}}\varphi_{n\mathbf{q}}^{\sigma}|\varphi_{n'\mathbf{q+k}}^{\sigma}\rangle\nonumber \end{eqnarray}
with the projections\begin{equation}
\langle\tilde{M}_{I}^{\mathbf{k}}\varphi_{n\mathbf{q}}^{\sigma}|\varphi_{n'\mathbf{q}+\mathbf{k}}^{\sigma}\rangle=\int\tilde{M}_{I}^{\mathbf{k}^{{\scriptstyle *}}}(\mathbf{r})\,\varphi_{n\mathbf{q}}^{\sigma^{{\scriptstyle *}}}(\mathbf{r})\,\varphi_{n'\mathbf{q}+\mathbf{k}}^{\sigma}(\mathbf{r})\, d^{3}r\label{eq:proj_MB}\end{equation}
and the Coulomb matrix \cite{Friedrich2009}\begin{equation}
v_{IJ}(\mathbf{k})=\langle M_{I}^{\mathbf{k}}|v|M_{J}^{\mathbf{k}}\rangle=\iint\frac{M_{I}^{\mathbf{k}^{{\scriptstyle *}}}(\mathbf{r})\, M_{J}^{\mathbf{k}}(\mathbf{r}')}{|\mathbf{r}-\mathbf{r}'|}\, d^{3}r\, d^{3}r'\,.\label{eq:CoulMat}\end{equation}
The sum over the occupied states also comprises the core states, which
give an important contribution to the exchange self-energy. Its evaluation
is simplified considerably by the fact that the core states can be
treated as dispersionless bands. We use a formalism derived by Dagens
and Perrot.\cite{Dagens1972} An efficient scheme for the calculation
of the full nonlocal Fock exchange potential including off-diagonal
elements will be presented elsewhere.\cite{Betzinger}

With the projections {[}Eq.~(\ref{eq:proj_MB}){]} we readily obtain
the representation of the polarization function,

\begin{eqnarray}
\lefteqn{P_{IJ}(\mathbf{k},\omega)}\label{eq:polar_MB}\\
 & = & \sum_{\sigma}\sum_{\mathbf{q}}^{\mathrm{BZ}}\sum_{n}^{\mathrm{occ.}}\sum_{n'}^{\mathrm{unocc}.}\langle\tilde{M}_{I}^{\mathbf{k}}\varphi_{n\mathbf{q}}^{\sigma}|\varphi_{n'\mathbf{q}+\mathbf{k}}^{\sigma}\rangle\langle\varphi_{n'\mathbf{q}+\mathbf{k}}^{\sigma}|\varphi_{n\mathbf{q}}^{\sigma}\tilde{M}_{J}^{\mathbf{k}}\rangle\nonumber \\
 & \times & \left(\frac{1}{\omega+\epsilon_{n\mathbf{q}}^{\sigma}-\epsilon_{n'\mathbf{q}+\mathbf{k}}^{\sigma}+i\eta}-\frac{1}{\omega-\epsilon_{n\mathbf{q}}^{\sigma}+\epsilon_{n'\mathbf{q}+\mathbf{k}}^{\sigma}-i\eta}\right)\,.\nonumber \end{eqnarray}
The rational expression in the brackets complicates a direct summation
over the Brillouin zone. It is more convenient to consider the representation
$(\textrm{Im}\, P)_{IJ}(\mathbf{k},\omega)$ of the imaginary part
$\textrm{Im}\, P(\mathbf{r},\mathbf{r}';\omega)$ first, which is
obtained by replacing expressions of the form $1/(a\pm i\eta)$ by
$\mp\pi\delta(a)$. This simplifies the BZ summation significantly.
Afterwards a Hilbert transformation yields the full polarization matrix
$P_{IJ}(\mathbf{k},\omega)$, where the frequency argument may be
complex. In particular, this allows an evaluation on the imaginary-frequency
axis, where the frequency-dependent quantities show a smooth behavior
and can therefore be sampled and interpolated with few frequency points.
As the bracket in Eq.~(\ref{eq:polar_MB}) is real for frequencies
on the imaginary axis, the corresponding matrix $P_{IJ}(\mathbf{k},i\omega)$
with $\omega\in\mathbb{R}$ becomes Hermitian; it even becomes real
symmetric if the system exhibits inversion symmetry and we use a symmetrized
MPB as described in Sec.~\ref{sec:mixedbasis}.

In the MPB, the integral equations for the dielectric function {[}Eq.~(\ref{eq:dielec}){]}
and the screened interaction {[}Eq.~(\ref{eq:screen_eps}){]} turn
into simple matrix equations. The equations become particularly simple
if we perform a basis transformation $\{ M_{I}^{\mathbf{k}}(\mathbf{r})\}\rightarrow\{ E_{\mu}^{\mathbf{k}}(\mathbf{r})\}$
that diagonalizes the Coulomb matrix. We note that no approximation
is involved at this stage. The new normalized basis functions are
necessarily orthogonal, and we do not need a biorthogonal set. In
this new basis the matrix equations become simple products,\begin{eqnarray}
\varepsilon_{\mu\nu}(\mathbf{k},\omega) & = & \delta_{\mu\nu}-\sqrt{v_{\mu}(\mathbf{k})}P_{\mu\nu}(\mathbf{k},\omega)\sqrt{v_{\nu}(\mathbf{k})}\label{eq:dielec_MB}\\
W_{\mu\nu}(\mathbf{k},\omega) & = & \sqrt{v_{\mu}(\mathbf{k})}\varepsilon_{\mu\nu}^{-1}(\mathbf{k},\omega)\sqrt{v_{\nu}(\mathbf{k})}\label{eq:screen_MB}\end{eqnarray}
with the eigenvalues $v_{\mu}(\mathbf{k})$ of the Coulomb matrix
{[}Eq.~(\ref{eq:CoulMat}){]}. Here we use a symmetrized definition
of the dielectric matrix $\varepsilon_{\mu\nu}(\mathbf{k},\omega)$
that is Hermitian (or real symmetric in case of inversion symmetry)
for imaginary frequencies and remains finite at the $\Gamma$ point.
It is easy to verify that the screened interaction remains unchanged
by this symmetrized formulation.

In contrast to the exchange self-energy, the frequency integral in
Eq.~(\ref{eq:selfenecorr_r}) cannot be replaced by a sum over residues
because the positions of the poles of $W_{\mu\nu}^{\mathrm{c}}(\mathbf{k},\omega)=W_{\mu\nu}(\mathbf{k},\omega)-\delta_{\mu\nu}v_{\mu}(\mathbf{k})$
in the complex-frequency plane are unknown. Therefore, the correlation
self-energy\begin{eqnarray}
\lefteqn{\langle\varphi_{n\mathbf{q}}^{\sigma}|\Sigma_{\mathrm{c}}^{\sigma}(\omega)|\varphi_{n\mathbf{q}}^{\sigma}\rangle}\label{eq:SEc_MB}\\
 & = & \frac{i}{2\pi}\sum_{\mathbf{k}}^{\mathrm{BZ}}\sum_{n'}^{\mathrm{all}}\sum_{\mu,\nu}\langle\varphi_{n'\mathbf{q+k}}^{\sigma}|\varphi_{n\mathbf{q}}^{\sigma}E_{\mu}^{\mathbf{k}}\rangle\langle E_{\nu}^{\mathbf{k}}\varphi_{n\mathbf{q}}^{\sigma}|\varphi_{n'\mathbf{q+k}}^{\sigma}\rangle\nonumber \\
 &  & \times\int_{-\infty}^{\infty}d\omega'\frac{W_{\mu\nu}^{\mathrm{c}}(\mathbf{k},\omega')}{\omega+\omega'-\epsilon_{n'\mathbf{q+k}}^{\sigma}+i\eta\,\textrm{sgn}(\epsilon_{n'\mathbf{q+k}}^{\sigma})}\nonumber \end{eqnarray}
still contains an explicit integration over frequencies. Unfortunately,
the integrand has a lot of structure along the real frequency axis,
which makes a direct evaluation difficult. There are two methods that
avoid the integration over real frequencies and use the imaginary
axis instead: analytic continuation \cite{Rojas1995} and contour
integration. \cite{Godby1988} The former allows a faster and easier
implementation, but contains a badly controlled fitting procedure,
which can be tested with the more accurate contour-integration method.
We have implemented both algorithms and find that they give similar
results for the systems considered here. In the following, we hence
focus exclusively on the first approach, which is based on an analytic
continuation of Eq.~(\ref{eq:SEc_MB}) to the imaginary-frequency
axis,\begin{eqnarray}
\lefteqn{\langle\varphi_{n\mathbf{q}}^{\sigma}|\Sigma_{\mathrm{c}}^{\sigma}(i\omega)|\varphi_{n\mathbf{q}}^{\sigma}\rangle}\label{eq:SEc_AC}\\
 & = & -\frac{1}{2\pi}\sum_{\mathbf{k}}^{\mathrm{BZ}}\sum_{n'}^{\mathrm{all}}\sum_{\mu,\nu}\langle\varphi_{n'\mathbf{q+k}}^{\sigma}|\varphi_{n\mathbf{q}}^{\sigma}E_{\mu}^{\mathbf{k}}\rangle\langle E_{\nu}^{\mathbf{k}}\varphi_{n\mathbf{q}}^{\sigma}|\varphi_{n'\mathbf{q+k}}^{\sigma}\rangle\nonumber \\
 &  & \times\int_{-\infty}^{\infty}d\omega'\frac{W_{\mu\nu}^{\mathrm{c}}(\mathbf{k},i\omega')}{i\omega+i\omega'-\epsilon_{n'\mathbf{q+k}}^{\sigma}}\,.\nonumber \end{eqnarray}
The integration contour can be closed over the positive imaginary
and negative real half-plane in Eqs.~(\ref{eq:SEc_MB}) and (\ref{eq:SEc_AC}),
respectively, and encloses the same poles. Now the frequency integration
is along the imaginary-frequency axis, where the integrand is much
smoother. In practice, we use a discrete and finite mesh for the imaginary
frequencies, which is dense near $\omega=0$. A tail is fitted to
the last mesh point according to the known asymptotic $\omega^{-2}$
behavior of the screened interaction. Between the mesh points, we
interpolate $W_{\mu\nu}^{\mathrm{c}}(\mathbf{k},\omega(\lambda))$
with cubic splines, where $\omega(\lambda)=\lambda/(1-\lambda)$ maps
the interval $[0,1[$ to $[0,\infty[$. This allows a stepwise analytic
integration. With this procedure only a small number of mesh points
is needed, typically around 10.

Once the correlation self-energy is calculated on the discrete imaginary-frequency
mesh, we analytically continue it to the whole complex plane by fitting
the model function,\begin{equation}
f(\omega)=\sum_{p=1}^{N_{\mathrm{p}}}\frac{a_{p}}{\omega-\omega_{p}}\quad\textrm{for\quad Re}\,\omega\ge0\label{eq:modelf}\end{equation}
with complex fit parameters $a_{p}$ and $\omega_{p}$. Due to the
location of the poles of the correlation self-energy in the complex
plane -- above the real axis for $\textrm{Re}\,\omega<0$ and below
the real axis for $\textrm{Re}\,\omega>0$ -- one must analytically
continue from the positive imaginary axis to the positive real axis.
For symmetry reasons one then obtains\begin{equation}
f(\omega)=\sum_{p=1}^{N_{\mathrm{p}}}\frac{a_{p}^{*}}{\omega-\omega_{p}^{*}}\quad\textrm{for\quad Re}\,\omega<0\end{equation}
on the negative real axis. In principle, $N_{\mathrm{p}}$ is a convergence
parameter. However, as the number of imaginary frequencies where $\langle\varphi_{n\mathbf{q}}^{\sigma}|\Sigma_{\mathrm{c}}^{\sigma}(i\omega)|\varphi_{n\mathbf{q}}^{\sigma}\rangle$
is known is relatively small, and as the fitting procedure quickly
becomes prohibitive for large numbers of fit parameters, one usually
uses only few poles, e.g., $N_{\mathrm{p}}=3$. After finding the
parameters $a_{p}$ and $\omega_{p}$, the correlation self-energy
$\langle\varphi_{n\mathbf{q}}^{\sigma}|\Sigma_{\mathrm{c}}^{\sigma}(\omega)|\varphi_{n\mathbf{q}}^{\sigma}\rangle$
is approximated by the analytic function $f(\omega)$, which allows
to solve the nonlinear quasiparticle equation (\ref{eq:quasipart_corr})
to machine precision with the standard iterative Newton method and
without any additional linearization of the self-energy.

\subsection{Brillouin-zone sampling}

Both the polarization function and the self-energy are defined as
products in real space. These become the convolutions {[}Eqs.~(\ref{eq:HFterm_MB}),
(\ref{eq:polar_MB}), and (\ref{eq:SEc_MB}){]} in a reciprocal-space
formulation, which is better suited for infinite periodic systems
because all nonlocal quantities then become block diagonal. We employ
the tetrahedron method for summations over the BZ. \cite{Rath1975}

These equations do not only contain the two Bloch vectors $\mathbf{k}$
and $\mathbf{q}$, but also their sum $\mathbf{k+q}$, at which the
KS wave functions and energies must be known. Therefore, we choose
the set of $\mathbf{k}$ points $\mathbf{k}_{n_{1}n_{2}n_{3}}=\sum_{i=1}^{3}n_{i}\mathbf{b}_{i}/N_{i}$
with $n_{i}=0,...,N_{i}-1$ and the reciprocal basis vectors $\mathbf{b}_{i}$.
We denote the number of $\mathbf{k}$ points by $N_{\mathbf{k}}=N_{1}N_{2}N_{3}$.
It naturally includes the point $\mathbf{k=0}$, which is special
because the long-range nature of the Coulomb interaction makes the
Coulomb matrix $v_{IJ}(\mathbf{k})$ and also the screened interaction
$W_{\mu\nu}(\mathbf{k},\omega)$ diverge in the limit $\mathbf{k}\rightarrow\mathbf{0}$.
This divergence must be taken into account in order to obtain fast
convergence with respect to the $\mathbf{k}$-point sampling. We will
discuss a numerically stable and efficient treatment in the next section.

\subsection{$\mathbf{\Gamma}$-point treatment}

The exchange and correlation self-energy contributions in Eqs.~(\ref{eq:HFterm_MB})
and (\ref{eq:SEc_AC}) each contain a sum over the BZ. The interaction
potentials $v_{IJ}(\mathbf{k})$ and $W_{\mu\nu}(\mathbf{k},\omega)$
diverge in the limit $\mathbf{k}\rightarrow\mathbf{0}$ but as this
pole is only of second order ($1/k^{2}$), a proper three-dimensional
integration over $\mathbf{k}=\mathbf{0}$ will yield a finite value. 

Likewise, the calculation of the dielectric matrix {[}Eq.~(\ref{eq:dielec_MB}){]}
involves a product of the polarization function and the divergent
Coulomb matrix. However, a closer inspection of the polarization matrix
{[}Eq.~(\ref{eq:polar_MB}){]} in the basis $\{ E_{\mu}^{\mathbf{k}}(\mathbf{r})\}$
shows that the head element $P_{11}(\mathbf{k},\omega)$ and the wing
elements $P_{\mu1}(\mathbf{k},\omega)$ and $P_{1\mu}(\mathbf{k},\omega)$
with $\mu>1$ are of the order $k^{2}$ and $k$, respectively, so
that the dielectric matrix $\varepsilon(\mathbf{k},\omega)$ remains
finite but angular-dependent at $\mathbf{k=0}$. \cite{Baroni1986}

In any case, the divergence gives an important contribution to the
self-energies and response functions and must be treated with care.
There are several numerical approaches. Kotani and van Schilfgaarde
\cite{Kotani2002} replaced the $\Gamma$ point by three additional
points $\mathbf{k}_{0}$ nearby, the so-called offset $\Gamma$ points.
These might be reduced by symmetry, e.g., to a single point in the
case of cubic symmetry. However, because of the summations in Eqs.~(\ref{eq:HFterm_MB}),
(\ref{eq:polar_MB}), and (\ref{eq:SEc_AC}), each additional point
$\mathbf{k}_{0}$ requires a complete auxiliary mesh $\{\mathbf{k}_{0}+\mathbf{q},\,\mathbf{q}\in\mathrm{BZ}\}$
on which the KS Hamiltonian must be diagonalized and the resulting
wave functions and energies must be stored. This at least doubles
the $\mathbf{k}$-point set, thus increasing the numerical cost in
terms of computation time and memory demand. In another approach,
Ku and Eguiluz \cite{Ku2002} as well as Puschnig and Ambrosch-Draxl
\cite{Puschnig2002} used a plane-wave basis for the Coulomb potential
and related propagators and thereby departed from a complete all-electron
description because plane waves are too inflexible to resolve the
rapid variations in the wave functions close to the atomic cores without
a prohibitively large basis set.

Here we present a scheme that does not require additional $\mathbf{k}$
points or projections onto plane waves. It thus combines the accuracy
of an all-electron approach with the numerical efficiency of a minimal
$\mathbf{k}$-point set. In the following, we present the algorithm
for the two self-energy contributions and the dielectric matrix.

\subsubsection{Exchange self-energy\label{sub:Exchange-self-energy}}

If the $\mathbf{k}$-point summation in Eq.~(\ref{eq:HFterm_MB})
is replaced by an integral, we can smoothly integrate over the divergence
of $v_{IJ}(\mathbf{k})$ and obtain a finite value. To this end, we
formally consider the Fourier transform\begin{eqnarray}
v_{\mathbf{GG}'}(\mathbf{k}) & = & \frac{1}{V}\int\frac{e^{-i(\mathbf{k}+\mathbf{G})\cdot\mathbf{r}}e^{i(\mathbf{k}+\mathbf{G}')\cdot\mathbf{r}'}}{|\mathbf{r}-\mathbf{r}'|}\, d^{3}r\, d^{3}r'\nonumber \\
 & = & \frac{4\pi}{k^{2}}\delta_{\mathbf{G0}}\delta_{\mathbf{G}'\mathbf{0}}+v_{\mathbf{GG}'}^{\prime}(\mathbf{k})\label{eq:v_fourier}\end{eqnarray}
with the nondivergent and infinitely large matrix $v_{\mathbf{GG}'}^{\prime}(\mathbf{k})=(1-\delta_{\mathbf{G0}})\delta_{\mathbf{GG}'}4\pi/|\mathbf{k+G}|^{2}$.
In a previous publication,\cite{Friedrich2009} we found an analogous
exact decomposition of the Coulomb matrix {[}Eq.~(\ref{eq:CoulMat}){]}
into the same divergent term and a nondivergent remainder $v_{IJ}^{\prime}(\mathbf{0})$.
Thus no projection onto plane waves is necessary, and we retain the
full accuracy of our all-electron formulation. Replacing $v_{\mathbf{GG}'}^{\prime}(\mathbf{0})$
by $v_{IJ}^{\prime}(\mathbf{0})$ and inserting Eq.~(\ref{eq:v_fourier})
into Eq.~(\ref{eq:HFterm_MB}) leads to contributions from the divergent
term \begin{equation}
\langle\varphi_{n\mathbf{q}}^{\sigma}|\Sigma_{\mathrm{x}}^{\sigma}|\varphi_{n\mathbf{q}}^{\sigma}\rangle_{\mathrm{div}}=-\frac{4\pi f_{n\mathbf{q}}^{\sigma}}{V}\left(\frac{V}{8\pi^{3}}\int_{\mathrm{BZ}}\frac{1}{k^{2}}\, d^{3}k-\sum_{\mathbf{k}\neq\mathbf{0}}^{\mathrm{BZ}}\frac{1}{k^{2}}\right)\label{eq:exchange_div}\end{equation}
with the occupation numbers $f_{n\mathbf{q}}^{\sigma}$ and from the
nondivergent term \begin{eqnarray}
\lefteqn{\langle\varphi_{n\mathbf{q}}^{\sigma}|\Sigma_{\mathrm{x}}^{\sigma}|\varphi_{n\mathbf{q}}^{\sigma}\rangle_{\mathrm{ndiv}}}\label{eq:exchange_ndiv}\\
 & = & -\sum_{\mathbf{k}}^{\mathrm{BZ}}\sum_{n'}^{\mathrm{occ.}}\sum_{I,J}v_{IJ}(\mathbf{k})\langle\tilde{M}_{J}^{\mathbf{k}}\varphi_{n\mathbf{q}}^{\sigma}|\varphi_{n'\mathbf{q+k}}^{\sigma}\rangle\nonumber \\
 &  & \quad\quad\quad\quad\quad\quad\quad\,\,\,\,\,\times\langle\varphi_{n'\mathbf{q+k}}^{\sigma}|\varphi_{n\mathbf{q}}^{\sigma}\tilde{M}_{I}^{\mathbf{k}}\rangle\,,\nonumber \end{eqnarray}
where we have set $v_{IJ}^{\prime}(\mathbf{0})\rightarrow v_{IJ}(\mathbf{0})$
for simplicity. The divergence of the Coulomb matrix is restricted
to the first term of Eq.~(\ref{eq:v_fourier}), and the corresponding
eigenfunction is $e^{i\mathbf{k}\cdot\mathbf{r}}/\sqrt{V}$, whose
$\mathbf{k}\rightarrow\mathbf{0}$ limit, a constant function, can
be represented exactly by the MPB\@. The products of $1/k^{2}$ with
higher-order terms of the projections $\langle e^{i\mathbf{k}\cdot\mathbf{r}}\varphi_{n\mathbf{q}}^{\sigma}|\varphi_{n'\mathbf{q+k}}^{\sigma}\rangle/\sqrt{V}$
can be of zeroth order and then lead to additional contributions to
Eq.~(\ref{eq:exchange_ndiv}) for $\mathbf{k=0}$. Therefore, the
projections must be expanded with the help of $\mathbf{k}\cdot\mathbf{p}$
perturbation theory. We have found that these corrections improve
the $\mathbf{k}$-point convergence significantly in some cases, while
in others -- especially for small band-gap semiconductors like GaAs
-- they worsen the convergence. The behavior also depends on the particular
electronic state $|\varphi_{n\mathbf{q}}^{\sigma}\rangle$. For simplicity,
we defer an in-depth discussion to a later publication.

In order to be able to integrate analytically, we extend the BZ integral
in Eq.~(\ref{eq:exchange_div}) to the whole reciprocal space by
replacing $1/k^{2}$ with\begin{equation}
F(\mathbf{k})=\sum_{\mathbf{G}}\frac{e^{-\beta|\mathbf{k+G}|^{2}}}{|\mathbf{k+G}|^{2}}\,.\end{equation}
Note that $F(\mathbf{k})$ diverges at every reciprocal lattice vector
$\mathbf{G}$. The exponential function was included to ensure the
convergence of the sum everywhere else. This function is formally
identical to the one used by Massidda \emph{et al.}~in Ref.~\onlinecite{Massidda1993}.
However, these authors define $\beta$ as a parameter depending on
the BZ size. Instead, we choose $\beta$ to be as small as possible
(while still allowing a sufficiently fast converging sum over $\mathbf{G}$),
independently of the BZ\@. First, this ensures that supercell calculations
of the same system yield identical values. Second, zero-order terms
arising from products of the exponential function with $1/k^{2}$
are small and can be neglected. After replacing $1/k^{2}$ with $F(\mathbf{k})$
and extending the integral and summation over the whole reciprocal
space we obtain \begin{eqnarray}
\lefteqn{\langle\varphi_{n\mathbf{q}}^{\sigma}|\Sigma_{\mathrm{x}}^{\sigma}|\varphi_{n\mathbf{q}}^{\sigma}\rangle_{\mathrm{div}}}\\
 & = & -4\pi f_{n\mathbf{q}}^{\sigma}\left(\frac{1}{8\pi^{3}}\int\frac{e^{-\beta q^{2}}}{q^{2}}\, d^{3}q-\frac{1}{\Omega N_{\mathbf{k}}}\sum_{0<|\mathbf{q}|}\frac{e^{-\beta q^{2}}}{q^{2}}\right)\,,\nonumber \end{eqnarray}
where the sum runs over all vectors $\mathbf{q=k+G\neq0}$ and $\Omega$
is the unit-cell volume. As both terms diverge for $\beta\rightarrow0$,
we introduce a cutoff radius $q_{0}$ and finally obtain\begin{eqnarray}
\lefteqn{\langle\varphi_{n\mathbf{q}}^{\sigma}|\Sigma_{\mathrm{x}}^{\sigma}|\varphi_{n\mathbf{q}}^{\sigma}\rangle_{\mathrm{div}}}\\
 & = & -f_{n\mathbf{q}}^{\sigma}\left[\frac{1}{\sqrt{\pi\beta}}\textrm{erf}(\sqrt{\beta}q_{0})-\frac{4\pi}{\Omega N_{\mathbf{k}}}\sum_{0<|\mathbf{q}|<q_{0}}\frac{e^{-\beta q^{2}}}{q^{2}}\right]\,.\nonumber \end{eqnarray}
We get rid of $q_{0}$ as a convergence parameter by choosing $e^{-\beta q_{0}^{2}}=\beta$
as a cutoff criterion for the summation. In practice, we find that
$\beta=0.005$ is typically small enough.

\subsubsection{Response functions\label{sub:Polar}}

The treatment of the polarization and dielectric function in the limit
$\mathbf{k\rightarrow0}$ is simplified considerably by the basis
transformation $\{ M_{I}^{\mathbf{k}}(\mathbf{r})\}\rightarrow\{ E_{\mu}^{\mathbf{k}}(\mathbf{r})\}$
introduced in Sec.~\ref{sub:Formulation-MPB} because it confines
the divergence of the Coulomb matrix to a single eigenvalue $v_{1}(\mathbf{k})\sim4\pi/k^{2}$
(Ref.~\onlinecite{Friedrich2009}). The corresponding eigenfunction
$E_{1}^{\mathbf{k}}(\mathbf{r})\sim e^{i\mathbf{k}\cdot\mathbf{r}}/\sqrt{V}$
is known analytically. 

Let us consider the projection {[}Eq.~(\ref{eq:proj_MB}){]}, where
the MPB function is replaced by this eigenfunction. Because of the
orthogonality of the wave functions, we have $\langle e^{i\mathbf{k}\cdot\mathbf{r}}\varphi_{n\mathbf{q}}^{\sigma}|\varphi_{n'\mathbf{q}+\mathbf{k}}^{\sigma}\rangle\sim\delta_{nn'}$
in the limit $\mathbf{k}\rightarrow\mathbf{0}$. For the moment, we
restrict ourselves to the case of semiconductors and insulators where
the band indices $n$ for occupied and $n'$ for unoccupied states
always differ such that in leading order $\langle e^{i\mathbf{k}\cdot\mathbf{r}}\varphi_{n\mathbf{q}}^{\sigma}|\varphi_{n'\mathbf{q}+\mathbf{k}}^{\sigma}\rangle\sim k$.
The linear order in $k$ exactly cancels the singularity of the Coulomb
matrix in the dielectric function {[}Eq.~(\ref{eq:dielec_MB}){]}
and can be calculated with $\mathbf{k}\cdot\mathbf{p}$ perturbation
theory,\cite{Baroni1986} which allows a full treatment of the divergence
and the dielectric anisotropy at $\mathbf{k=0}$. \cite{Freysoldt2007}
In this way, the matrices decompose into head, wings, and body as
in a simple plane-wave basis set. Still, the all-electron accuracy
is fully retained and no projection onto plane waves is necessary.

For the sake of completeness, here we give the exact expressions for
the polarization function and the screened interaction, taking into
account the full anisotropy. From $\mathbf{k}\cdot\mathbf{p}$ perturbation
theory, one obtains the form \begin{equation}
P_{\mu\nu}(\mathbf{k},i\omega)\sim\left(\begin{array}{cccc}
\mathbf{e}_{\mathbf{k}}^{\mathrm{T}}\mathbf{H}(\omega)\mathbf{e_{k}}k^{2} & \mathbf{e}_{\mathbf{k}}^{\mathrm{T}}\mathbf{s}_{2}(\omega)k & \cdots & \mathbf{e}_{\mathbf{k}}^{\mathrm{T}}\mathbf{s}_{n}(\omega)k\\
\mathbf{e}_{\mathbf{k}}^{\mathrm{T}}\mathbf{s}_{2}^{*}(\omega)k & \tilde{P}_{22}(\omega) & \cdots & \tilde{P}_{2n}(\omega)\\
\vdots & \vdots & \ddots & \vdots\\
\mathbf{e}_{\mathbf{k}}^{\mathrm{T}}\mathbf{s}_{n}^{*}(\omega)k & \tilde{P}_{n2}(\omega) & \cdots & \tilde{P}_{nn}(\omega)\end{array}\right)\label{eq:polar_Gamma}\end{equation}
for the polarization function in the limit $\mathbf{k}\rightarrow\mathbf{0}$.
Here $\mathbf{H}(\omega)$ is a $3\times3$ matrix, $\mathbf{s}_{\mu}(\omega)$
are three-dimensional vectors, and the matrix elements $\tilde{P}_{\mu\nu}(\omega)$
are finite. We note that for frequency arguments that are not purely
imaginary the matrix $P_{\mu\nu}(\mathbf{k},\omega)$ is not Hermitian;
in particular, the horizontal and vertical wings are then not simply
the complex conjugates of one another. Otherwise, the formalism is
very similar to the one given here. The corresponding screened interaction
becomes\begin{widetext}\begin{eqnarray}
W_{\mu\nu}(\mathbf{k},i\omega) & \sim & \left(\begin{array}{cccc}
0 & 0 & \cdots & 0\\
0 & \tilde{W}_{22}(\omega) & \cdots & \tilde{W}_{2n}(\omega)\\
\vdots & \vdots & \ddots & \vdots\\
0 & \tilde{W}_{n2}(\omega) & \cdots & \tilde{W}_{nn}(\omega)\end{array}\right)\label{eq:screen_Gamma}\\
 &  & +\frac{1}{\mathbf{e}_{\mathbf{k}}^{\mathrm{T}}\mathbf{L}(\omega)\mathbf{e_{k}}}\left(\begin{array}{cccc}
4\pi/k^{2} & \mathbf{e}_{\mathbf{k}}^{\mathrm{T}}\mathbf{w}_{2}(\omega)/k & \cdots & \mathbf{e}_{\mathbf{k}}^{\mathrm{T}}\mathbf{w}_{n}(\omega)/k\\
\mathbf{e}_{\mathbf{k}}^{\mathrm{T}}\mathbf{w}_{2}^{*}(\omega)/k & \left|\mathbf{e}_{\mathbf{k}}^{\mathrm{T}}\mathbf{w}_{2}(\omega)\right|^{2} & \cdots & \left[\mathbf{e}_{\mathbf{k}}^{\mathrm{T}}\mathbf{w}_{2}^{*}(\omega)\right]\left[\mathbf{e}_{\mathbf{k}}^{\mathrm{T}}\mathbf{w}_{n}(\omega)\right]\\
\vdots & \vdots & \ddots & \vdots\\
\mathbf{e}_{\mathbf{k}}^{\mathrm{T}}\mathbf{w}_{n}^{*}(\omega)/k & \left[\mathbf{e}_{\mathbf{k}}^{\mathrm{T}}\mathbf{w}_{n}^{*}(\omega)\right]\left[\mathbf{e}_{\mathbf{k}}^{\mathrm{T}}\mathbf{w}_{2}(\omega)\right] & \cdots & \left|\mathbf{e}_{\mathbf{k}}^{\mathrm{T}}\mathbf{w}_{n}(\omega)\right|^{2}\end{array}\right)\nonumber \end{eqnarray}
\end{widetext} with the finite matrix elements \begin{eqnarray}
\tilde{W}_{\mu\nu}(\omega) & = & \sqrt{v_{\mu}(\mathbf{0})}\tilde{\varepsilon}_{\mu\nu}^{-1}(\omega)\sqrt{v_{\nu}(\mathbf{0})}\,,\\
\tilde{\varepsilon}_{\mu\nu}(\omega) & = & \delta_{\mu\nu}-\sqrt{v_{\mu}(\mathbf{0})}\tilde{P}_{\mu\nu}(\omega)\sqrt{v_{\nu}(\mathbf{0})}\,,\end{eqnarray}
where $\mu,\nu\ge2$. The divergent and, in general, angular-dependent
second term derives from the head and wing elements of Eq.~(\ref{eq:polar_Gamma})
with

\begin{equation}
\mathbf{w}_{\mu}(\omega)=\sqrt{4\pi}\sum_{\nu\ge2}\tilde{\epsilon}_{\mu\nu}^{-1}(\omega)\mathbf{s}_{\nu}(\omega)\sqrt{v_{\nu}(\mathbf{0})}\end{equation}
and\begin{equation}
\mathbf{L}(\omega)=\mathbf{1}-4\pi\mathbf{H}(\omega)-\sqrt{4\pi}\sum_{\mu\ge2}\mathbf{s}_{\mu}(\omega)\mathbf{w}_{\mu}^{\mathrm{T}}(\omega)\sqrt{v_{\mu}(\mathbf{0})}\,.\end{equation}

Let us now turn to the case of a metallic system, where in addition
to the interband transitions with $n\neq n'$, there is a contribution
from intraband transitions across the Fermi surface. These transitions
occur within one electron band, i.e., $n=n'$, in which case the projection
above becomes unity in the limit $\mathbf{k}\rightarrow\mathbf{0}$.
However, it can be shown\cite{Ziesche1983} that the expression in
the brackets of Eq.~(\ref{eq:polar_MB}) will then be of linear order
in $\mathbf{k}$ and that we obtain a contribution only for the head
element of the polarization matrix, the so-called Drude term\begin{equation}
P_{11}^{\mathrm{D}}(\mathbf{k},i\omega)\sim-\frac{k^{2}}{4\pi}\frac{\omega_{\mathrm{pl}}^{2}}{\omega(\omega+i\eta)}\,,\label{eq:PDrude}\end{equation}
where $\omega_{\mathrm{pl}}$ is the plasma frequency obtained by
an integration over the Fermi surface. The Drude term gives rise to
a contribution $\omega_{\mathrm{pl}}^{2}/[\omega(\omega+i\eta)]$
for the head element of the dielectric matrix {[}Eq.~(\ref{eq:dielec_MB}){]},
which will mix with all other elements in the inversion for the screened
interaction {[}Eq.~(\ref{eq:screen_MB}){]}. However, we find that
$W_{11}^{\mathrm{c}}(\mathbf{k},\omega)$ is dominated by the \emph{bare}
Drude term \begin{equation}
W_{11}^{\mathrm{c,D}}(\mathbf{k},i\omega)\sim-\frac{4\pi}{k^{2}}\frac{\omega_{\mathrm{pl}}^{2}}{\omega^{2}+\omega_{\mathrm{p}}^{2}}\end{equation}
 in the limit $\mathbf{k}\rightarrow\mathbf{0}$. As this expression
can be convoluted with the Green function in Eq.~(\ref{eq:SEc_AC})
analytically, we subtract it from the head element and treat the remainder
$W_{11}^{\mathrm{c}}(\mathbf{k},i\omega)-W_{11}^{\mathrm{c,D}}(\mathbf{k},i\omega)|_{\mathbf{k}\rightarrow\mathbf{0}}$
numerically. The treatment of the $1/k^{2}$ divergence is explained
in the next section.

\subsubsection{Correlation self-energy\label{sub:selfenecorr}}

The BZ summation over the $1/k^{2}$ divergence in the correlation
self-energy {[}Eq.~(\ref{eq:SEc_MB}){]} can be treated with the
same procedure as outlined in Sec.~\ref{sub:Exchange-self-energy}
for the exchange self-energy. However, in this case, there are additional
$1/k$ terms, and all divergent terms exhibit an additional angular
dependence. 

As a first step we describe this angular dependence with the help
of spherical harmonics. For example, for the head element we must
find the coefficients $K_{(2l)m}(\omega)$ in \begin{equation}
\frac{4\pi}{\mathbf{e}_{\mathbf{k}}^{\mathrm{T}}\mathbf{L}(\omega)\mathbf{e_{k}}}=\sum_{l=0}^{\infty}\sum_{m=-2l}^{2l}K_{(2l)m}(\omega)Y_{(2l)m}(\mathbf{e_{k}})\,.\end{equation}
If we multiply with the denominator $\mathbf{e}_{\mathbf{k}}^{\mathrm{T}}\mathbf{L}(\omega)\mathbf{e_{k}}=\sum_{l=0}^{1}\sum_{m=-l}^{l}L_{(2l)m}(\omega)Y_{(2l)m}^{*}(\mathbf{e_{k}})$,
use the Gaunt coefficients and the orthogonality of spherical harmonics,
we obtain an infinite system of linear equations, from which the coefficients
$K_{(2l)m}(\omega)$ can be deduced. The corresponding expansions
for the wing,\begin{equation}
\frac{\mathbf{e}_{\mathbf{k}}^{\mathrm{T}}\mathbf{w}_{\mu}(\omega)}{\mathbf{e}_{\mathbf{k}}^{\mathrm{T}}\mathbf{L}(\omega)\mathbf{e_{k}}}=\sum_{l=0}^{\infty}\sum_{m=-(2l+1)}^{2l+1}K_{\mu,(2l+1)m}(\omega)Y_{(2l+1)m}(\mathbf{e_{k}})\label{eq:wingang}\end{equation}
and body matrix elements can be calculated in a similar way. Finally,
subtraction of $\delta_{\mu\nu}v_{\mu}(\mathbf{0})$ with $v_{1}(\mathbf{0})\sim4\pi/k^{2}$
yields the expansions for the head and body of $W^{\mathrm{c}}$.
Then $K_{(2l)m}^{\mathrm{c}}(\omega)=K_{(2l)m}(\omega)-(4\pi)^{3/2}\delta_{l0}$
and $\tilde{W}_{\mu\nu}^{\mathrm{c}}(\omega)=\tilde{W}_{\mu\nu}(\omega)-\delta_{\mu\nu}v_{\mu}(\mathbf{0})$
replace $K_{(2l)m}(\omega)$ and $\tilde{W}_{\mu\nu}(\omega)$, respectively.

The body matrix elements of the second term in Eq.~(\ref{eq:screen_Gamma})
are angular-dependent but finite. Then all terms with $l>0$ integrate
to zero, and we retain only the constant term $l=0$, which we simply
add to $\tilde{W}_{\mu\nu}^{\mathrm{c}}$. The head (wing) elements
diverge with a factor $k^{-2}$ ($k^{-1}$). As in Sec.~\ref{sub:Exchange-self-energy},
multiplication with higher orders of the projections and the term
$1/(i\omega+i\omega'-\epsilon_{n'\mathbf{q+k}}^{\sigma})$ leads to
terms of zeroth order in $\mathbf{k}$. Again we do not discuss these
terms explicitly, as they improve the results only in some cases,
while in others, they can lead to numerical problems. This can be
attributed to the energy denominators of $\mathbf{k}\cdot\mathbf{p}$
perturbation theory, which is used to compute the higher-order terms. 

The head element exhibits a $1/k^{2}$ divergence, which can be treated
in the same way as in Sec.~\ref{sub:Exchange-self-energy}. We obtain
a contribution \begin{eqnarray}
\lefteqn{\langle\varphi_{n\mathbf{q}}^{\sigma}|\Sigma_{\mathrm{c}}^{\sigma}(i\omega)|\varphi_{n\mathbf{q}}^{\sigma}\rangle_{\mathrm{div}}}\label{eq:correlation_div}\\
 & = & -\frac{1}{4\pi^{3/2}V}\left(\frac{V}{8\pi^{3}}\int_{\mathrm{BZ}}\frac{1}{k^{2}}\, d^{3}k-\sum_{\mathbf{k}\neq\mathbf{0}}^{\mathrm{BZ}}\frac{1}{k^{2}}\right)\nonumber \\
 &  & \times\int_{-\infty}^{\infty}\frac{K_{00}^{\mathrm{c}}(\omega')}{i\omega+i\omega'-\epsilon_{n\mathbf{q}}^{\sigma}}d\omega'\,,\nonumber \end{eqnarray}
where the frequency integration is performed as described in Sec.~\ref{sub:Formulation-MPB}.
All other elements $K_{(2l)m}^{\mathrm{c}}(\omega)$ as well as the
divergent wing elements of Eq.~(\ref{eq:screen_Gamma}) need not
be taken into account, as their angular parts integrate to zero. However,
we note that there is a finite contribution of these elements from
multiplications with higher-order terms of the other quantities --
a contribution that we neglect here for simplicity, as previously
mentioned.

\subsection{Optimization of the MPB\label{sub:OptimizationMPB}}

If we assume that the eigenvalues $v_{\mu}(\mathbf{k})$ are ordered
according to decreasing size, then matrix elements $\varepsilon_{\mu\nu}(\mathbf{k},\omega)$
and $W_{\mu\nu}(\mathbf{k},\omega)$ with large indices will be relatively
small, cf.~Eqs.~(\ref{eq:dielec_MB}) and (\ref{eq:screen_MB}).
We may then introduce a threshold value $v_{\mathrm{min}}$ for the
eigenvalues and only retain the functions $E_{\mu}^{\mathbf{k}}(\mathbf{r})$
with $v_{\mu}(\mathbf{k})\ge v_{\mathrm{min}}$. As the eigenvalue
$v_{\mu}(\mathbf{k})$ can be viewed as a measure for the importance
of the corresponding function $E_{\mu}^{\mathbf{k}}(\mathbf{r})$
in $v(\mathbf{r})$, we restrict ourselves to the dominant part of
the electron-electron interaction in this way. The removal of basis
functions with small eigenvalues can be viewed as an optimization
step of the MPB, because it reduces the matrix sizes and hence the
computational cost considerably without compromising the accuracy,
as we show in Sec.~\ref{sec:Test} below. We there also demonstrate
that the results converge reasonably fast with respect to the threshold
parameter $v_{\mathrm{min}}$. Note that with $v_{\mathrm{min}}=0$,
the full accuracy of the MPB is restored. In our implementation, this
optimization of the MPB only affects the correlation self-energy while
we always calculate the exchange self-energy with the full MPB.

\subsection{Use of symmetry\label{sub:Symmetry}}

The evaluation of Eqs.~(\ref{eq:HFterm_MB}), (\ref{eq:polar_MB}),
and (\ref{eq:SEc_AC}) takes considerable computation time, which
can be reduced substantially by exploiting spatial and time-reversal
symmetries of the system, the latter in case of a system without inversion
symmetry. Allowed symmetry operations are those that leave the Hamiltonian
invariant. With these operations, the set of $\mathbf{k}$ vectors
decomposes into groups of equivalent vectors, which are equivalent
in the sense that all elements of the group can be generated by applying
the symmetry operations to an arbitrary representative of the group
elements. As a consequence, any physical quantity defined for the
representative $\mathbf{k}$ vector can be mapped to any other vector
of the group by a suitable symmetry operation. This reduces the full
BZ sampling to the smaller set of representative vectors, which form
the irreducible BZ (IBZ).

We may thus restrict $P_{IJ}(\mathbf{k},\omega)$ to $\mathbf{k}\in\mathrm{IBZ}$.
The summation over $\mathbf{q}$ points in Eq.~(\ref{eq:polar_MB}),
on the other hand, cannot be confined in the same way, because the
terms to be summed also depend on $\mathbf{k}$ (and $\mathbf{q+k}$).
However, we can restrict the $\mathbf{q}$ vectors to an \emph{extended}
IBZ {[}EIBZ($\mathbf{k}$){]} that is defined in the same way as the
IBZ above but with the subset of symmetry operations that leave the
given $\mathbf{k}$ vector invariant. 

Let us define the complete set of $N_{A}$ symmetry operations by\begin{equation}
S_{A}=\{\hat{{\cal A}}_{i}=(\mathbf{A}_{i},\mathbf{a}_{i},\alpha_{i})|i=1,...,N_{A}\}\,,\end{equation}
where $\mathbf{A}_{i}$ and $\mathbf{a}_{i}$ denote the $3\times3$
rotation (or rotoinversion) matrix and a translation vector (which
is nonzero for nonsymmorphic operations), respectively, and $\alpha_{i}$
equals $-1$ ($+1$) for operations with (without) time reversal.
The action of $\hat{A}_{i}$ on a spatial vector $\mathbf{r}$, a
momentum vector $\mathbf{k}\in\mathrm{BZ}$, and a function $f(\mathbf{r})$
is declared by\begin{eqnarray}
\hat{A}_{i}\mathbf{r} & = & \mathbf{A}_{i}\mathbf{r}+\mathbf{a}_{i}\,,\\
\hat{A}_{i}\mathbf{k} & = & \alpha_{i}\mathbf{A}_{i}\mathbf{k}+\overline{\mathbf{G}}\,,\end{eqnarray}
\begin{equation}
\hat{A}_{i}f(\mathbf{r})=\begin{cases}
f\left[\mathbf{A}^{-1}(\mathbf{r}-\mathbf{a}_{i})\right] & \textrm{ for }\alpha_{i}=1\\
f^{*}\left[\mathbf{A}^{-1}(\mathbf{r}-\mathbf{a}_{i})\right] & \textrm{ for }\alpha_{i}=-1\,,\end{cases}\end{equation}
where the reciprocal lattice vector $\overline{\mathbf{G}}$ folds
$\alpha_{i}\mathbf{A}_{i}\mathbf{k}$ back into the BZ. Furthermore,
we define the subset\begin{equation}
S_{A}^{\mathbf{k}}=\{\hat{A}_{i}^{\mathbf{k}}|i=1,...,N_{A}^{\mathbf{k}};\hat{A}_{i}^{\mathbf{k}}\mathbf{k}=\mathbf{\mathbf{k}}\}\in S_{A}\end{equation}
that generates the EIBZ($\mathbf{k}$).

Now we reformulate Eq.~(\ref{eq:polar_MB}) using the definition
of the EIBZ(\textbf{$\mathbf{k}$}) and that $\hat{A}_{i}\varphi_{n\mathbf{q}}^{\sigma}(\mathbf{r})$
is a valid wave function with the momentum vector $\hat{A}_{i}\mathbf{q}$,\begin{eqnarray}
\lefteqn{P_{IJ}(\mathbf{k},\omega)=\sum_{i=1}^{N_{A}^{\mathbf{k}}}\sum_{\sigma}\sum_{\mathbf{q}}^{\mathrm{EIBZ(\mathbf{k})}}\frac{N_{\mathbf{q}}^{\mathbf{k}}}{N_{A}^{\mathbf{k}}}}\label{eq:polar2}\\
 &  & \times\sum_{n}^{\mathrm{occ}}\sum_{n'}^{\mathrm{unocc}}\langle\tilde{M}_{I}^{\mathbf{k}}\hat{A}_{i}^{\mathbf{k}^{{\scriptstyle -1}}}\varphi_{n\mathbf{q}}^{\sigma}|\hat{A}_{i}^{\mathbf{k}^{{\scriptstyle -1}}}\varphi_{n'\mathbf{q+k}}^{\sigma}\rangle\nonumber \\
 &  & \quad\quad\quad\,\,\,\,\,\,\times\langle\hat{A}_{i}^{\mathbf{k}^{{\scriptstyle -1}}}\varphi_{n'\mathbf{q+k}}^{\sigma}|\hat{A}_{i}^{\mathbf{k}^{{\scriptstyle -1}}}\varphi_{n\mathbf{q}}^{\sigma}\tilde{M}_{J}^{\mathbf{k}}\rangle\left(...\right)\nonumber \\
 & = & \sum_{i=1}^{N_{A}^{\mathbf{k}}}\sum_{\sigma}\sum_{\mathbf{q}}^{\mathrm{EIBZ(\mathbf{k})}}\frac{N_{\mathbf{q}}^{\mathbf{k}}}{N_{A}^{\mathbf{k}}}\sum_{n}^{\mathrm{occ}}\sum_{n'}^{\mathrm{unocc}}\hat{T}_{\alpha_{i}}\left[\langle\hat{A}_{i}^{\mathbf{k}}\tilde{M}_{I}^{\mathbf{k}}\varphi_{n\mathbf{q}}^{\sigma}|\varphi_{n'\mathbf{q+k}}^{\sigma}\rangle\right.\nonumber \\
 &  & \quad\quad\quad\quad\quad\quad\quad\quad\quad\quad\quad\left.\times\langle\varphi_{n'\mathbf{q+k}}^{\sigma}|\varphi_{n\mathbf{q}}^{\sigma}\hat{A}_{i}^{\mathbf{k}}\tilde{M}_{J}^{\mathbf{k}}\rangle\right]\left(...\right)\nonumber \end{eqnarray}
where $N_{\mathbf{q}}^{\mathbf{k}}$ is the number of equivalent $\mathbf{q}$
vectors with respect to $S_{A}^{\mathbf{k}}$ , $\hat{T}_{1}$ the
identity, and $\hat{T}_{-1}$ the transpose operator $\hat{T}_{-1}B_{IJ}=B_{JI}$.
From the definition of the MPB, it is clear that the application of
an arbitrary symmetry operation $\hat{A}_{i}^{\mathbf{k}}\in S_{A}^{\mathbf{k}}$
to any $M_{I}^{\mathbf{k}}(\mathbf{r})$ can be written as a linear
combination of the basis functions $M_{J}^{\mathbf{k}}(\mathbf{r})$,
such that the sum over the symmetry operations in Eq.~(\ref{eq:polar2})
can be performed at the very end after summing over the bands, the
EIBZ($\mathbf{k}$), and the spins. We note that this is also possible
with the set $\{ E_{\mu}^{\mathbf{k}}(\mathbf{r})\}$ instead of $\{ M_{I}^{\mathbf{k}}(\mathbf{r})\}$.

In a similar way, we can accelerate the evaluation of the expectation
values of $\Sigma_{\mathrm{x}}^{\sigma}$ and $\Sigma_{\mathrm{c}}^{\sigma}(\omega)$.
To this end, we write Eqs.~(\ref{eq:HFterm_MB}) and (\ref{eq:SEc_AC})
in a common general form with a function $f(\mathbf{r},\mathbf{r}')$,
which fulfills all symmetry properties of the system. By confining
the summation over $\mathbf{k}$ points to the EIBZ($\mathbf{q}$)
and summing over the symmetry operations we obtain\begin{widetext}\begin{eqnarray}
\langle\varphi_{n\mathbf{q}}^{\sigma}|\Sigma^{\sigma}|\varphi_{n\mathbf{q}}^{\sigma}\rangle & = & \sum_{\mathbf{k}}^{\mathrm{BZ}}\sum_{n'}\iint d^{3}r\, d^{3}r'\,\varphi_{n\mathbf{q}}^{\sigma^{{\scriptstyle *}}}\left(\mathbf{r}\right)\varphi_{n'\mathbf{k}}^{\sigma}\left(\mathbf{r}\right)\varphi_{n'\mathbf{k}}^{\sigma^{{\scriptstyle *}}}\left(\mathbf{r}'\right)\varphi_{n\mathbf{q}}^{\sigma}\left(\mathbf{r}'\right)f(\mathbf{r},\mathbf{r}')\\
 & = & \sum_{i=1}^{N_{A}^{\mathbf{q}}}\sum_{\mathbf{k}}^{\mathrm{EIBZ}(\mathbf{q})}\frac{N_{\mathbf{k}}^{\mathbf{q}}}{N_{A}^{\mathbf{q}}}\sum_{n'}\iint d^{3}r\, d^{3}r'\,\varphi_{n\mathbf{q}}^{\sigma^{{\scriptstyle *}}}\left(\mathbf{r}\right)[\hat{A}_{i}^{\mathbf{q}^{{\scriptstyle -1}}}\varphi_{n'\mathbf{k}}^{\sigma}\left(\mathbf{r}\right)][\hat{A}_{i}^{\mathbf{q}^{{\scriptstyle -1}}}\varphi_{n'\mathbf{k}}^{\sigma^{{\scriptstyle *}}}\left(\mathbf{r}'\right)]\varphi_{n\mathbf{q}}^{\sigma}\left(\mathbf{r}'\right)f(\mathbf{r},\mathbf{r}')\nonumber \\
 & = & \sum_{i=1}^{N_{A}^{\mathbf{q}}}\sum_{\mathbf{k}}^{\mathrm{EIBZ}(\mathbf{q})}\frac{N_{\mathbf{k}}^{\mathbf{q}}}{N_{A}^{\mathbf{q}}}\sum_{n'}\iint d^{3}r\, d^{3}r'\,[\hat{A}_{i}^{\mathbf{q}}\varphi_{n\mathbf{q}}^{\sigma^{{\scriptstyle *}}}\left(\mathbf{r}\right)]\varphi_{n'\mathbf{k}}^{\sigma}\left(\mathbf{r}\right)\varphi_{n'\mathbf{k}}^{\sigma^{{\scriptstyle *}}}\left(\mathbf{r}'\right)[\hat{A}_{i}^{\mathbf{q}}\varphi_{n\mathbf{q}}^{\sigma}\left(\mathbf{r}'\right)]f(\mathbf{r},\mathbf{r}')\nonumber \\
 & = & \sum_{\mathbf{k}}^{\mathrm{EIBZ}(\mathbf{q})}\frac{N_{\mathbf{k}}^{\mathbf{q}}}{N_{A}^{\mathbf{q}}}\sum_{m}\sum_{m'}\left(\sum_{i=1}^{N_{A}^{\mathbf{q}}}A_{i,mn\mathbf{q}}^{\sigma^{{\scriptstyle *}}}A_{i,m'n\mathbf{q}}^{\sigma}\right)\sum_{n'}\iint d^{3}r\, d^{3}r'\,\varphi_{m\mathbf{q}}^{\sigma^{{\scriptstyle *}}}\left(\mathbf{r}\right)\varphi_{n'\mathbf{k}}^{\sigma}\left(\mathbf{r}\right)\varphi_{n'\mathbf{k}}^{\sigma^{{\scriptstyle *}}}\left(\mathbf{r}'\right)\varphi_{m'\mathbf{q}}^{\sigma}\left(\mathbf{r}'\right)f(\mathbf{r},\mathbf{r}')\,,\nonumber \end{eqnarray}
\end{widetext} where we have restricted ourselves for simplicity
to symmetry operations without time reversal. $A_{i,mn\mathbf{q}}^{\sigma}=\langle\varphi_{m\mathbf{q}}^{\sigma}|\hat{A}_{i}^{\mathbf{q}}|\varphi_{n\mathbf{q}}^{\sigma}\rangle$
is the matrix representation of $\hat{A}_{i}^{\mathbf{q}}$ in terms
of the wave functions. As $\hat{A}_{i}^{\mathbf{q}}$ commutes with
the Hamiltonian, the element $A_{i,mn\mathbf{q}}^{\sigma}$ can only
be nonzero if the corresponding energies $\epsilon_{m\mathbf{q}}^{\sigma}$
and $\epsilon_{n\mathbf{q}}^{\sigma}$ are identical. Let us assume
that $\varphi_{m\mathbf{q}}^{\sigma}$ and $\varphi_{n\mathbf{q}}^{\sigma}$
lie in the eigenspace formed by $\varphi_{\nu\mathbf{q}}^{\sigma}$
with $n_{1}\le\nu\le n_{2}$. By construction $A_{i,mn\mathbf{q}}^{\sigma}$
is then an irreducible representation, and we may apply the great
orthogonality theorem of group theory,\cite{Kim}\begin{equation}
\sum_{i=1}^{N_{A}^{\mathbf{q}}}A_{i,mn\mathbf{q}}^{\sigma^{{\scriptstyle *}}}A_{i,m'n\mathbf{q}}^{\sigma}=\frac{N_{A}^{\mathbf{q}}}{n_{2}-n_{1}+1}\delta_{mm'}\,,\end{equation}
which finally yields\begin{eqnarray}
\lefteqn{\langle\varphi_{n\mathbf{q}}^{\sigma}|\Sigma^{\sigma}|\varphi_{n\mathbf{q}}^{\sigma}\rangle=\sum_{\mathbf{k}}^{\mathrm{EIBZ}(\mathbf{q})}\frac{N_{\mathbf{k}}^{\mathbf{q}}}{n_{2}-n_{1}+1}}\label{eq:SelfEnergySym}\\
 &  & \times\sum_{m=n_{1}}^{n_{2}}\sum_{n'}\iint d^{3}r\, d^{3}r'\,\varphi_{m\mathbf{q}}^{\sigma^{{\scriptstyle *}}}\left(\mathbf{r}\right)\varphi_{n'\mathbf{k}}^{\sigma}\left(\mathbf{r}\right)\varphi_{n'\mathbf{k}}^{\sigma^{{\scriptstyle *}}}\left(\mathbf{r}'\right)\nonumber \\
 &  & \quad\quad\quad\quad\quad\quad\quad\quad\quad\quad\quad\quad\quad\quad\quad\times\varphi_{m\mathbf{q}}^{\sigma}\left(\mathbf{r}'\right)f(\mathbf{r},\mathbf{r}')\,.\nonumber \end{eqnarray}
The $\mathbf{k}$-point sum is thus reduced to the EIBZ($\mathbf{q}$),
but we have to average over the degenerate states $n_{1}\le m\le n_{2}$.
However, the gain in computation time by a restriction to the EIBZ(\textbf{$\mathbf{q}$})
usually outweighs the overhead from the summation over the degenerate
states by far. For symmetry groups with time-reversal symmetries the
derivation can be done analogously with a more general great orthogonality
theorem.\cite{Kim} The final result is identical to Eq.~(\ref{eq:SelfEnergySym}).

\section{Test calculations\label{sec:Test}}

We have implemented above algorithm in the computer program \noun{spex}.
In the following, we first show detailed convergence tests for Si
and $\mathrm{SrTiO}_{3}$. Silicon is a prototype semiconductor, for
which many $GW$ calculations already exist and which is therefore
used as a benchmark material. Strontium titanate is a prototype transition-metal
oxide, which crystallizes in the frequently occurring perovskite structure.
It is currently explored as a high-$\kappa$ dielectric and a promising
barrier material in spintronics and nanoelectronics. The valence and
the lowest conduction bands are formed by O $2p$ and Ti $3d$ states,
respectively. We explicitly include the semicore $3s$ and $3p$ states
of Ti as well as the $4s$ and $4p$ states of Sr with the help of
local orbitals and take their contribution to the screening into account.
All these states are accurately described by the FLAPW basis set.
Additional local orbitals are used to improve the description of high-lying
unoccupied states.\cite{Friedrich2006} As reference, we also show
an overview of $GW$ results for a wide range of semiconductors including
$\mathrm{BaTiO}_{3}$ and compare them with experimental and theoretical
values from the literature. Furthermore, the efficiency of our scheme
is illustrated by calculations for diamond supercells containing 16
and 128 atoms.

The numerical procedure involves a number of convergence parameters,
which determine the accuracy of convolutions in real space (MPB),
reciprocal space ($\mathbf{k}$-point set), and the frequency domain.
Since the latter two apply to essentially all electronic-structure
methods, we concentrate mainly on the parameters for the MPB here.
Specifically, we consider the cutoff parameters $L_{\mathrm{max}}$
for the angular momentum inside the MT spheres and $G'_{\mathrm{max}}$
for the IPWs as well as the threshold $v_{\mathrm{min}}$ for the
optimization of the MPB according to Sec.~\ref{sub:OptimizationMPB}.
We also discuss the convergence with respect to the number of unoccupied
states for the summations in Eqs.~(\ref{eq:polar_MB}) and (\ref{eq:SEc_AC}).
All calculations are done with a 4$\times$4$\times$4 $\mathbf{k}$-point
set and the local-density approximation (LDA) (Ref.~\onlinecite{Perdew1981})for
the exchange-correlation functional at the DFT level.

\begin{figure}
\includegraphics[%
  width=0.70\columnwidth,
  angle=-90]{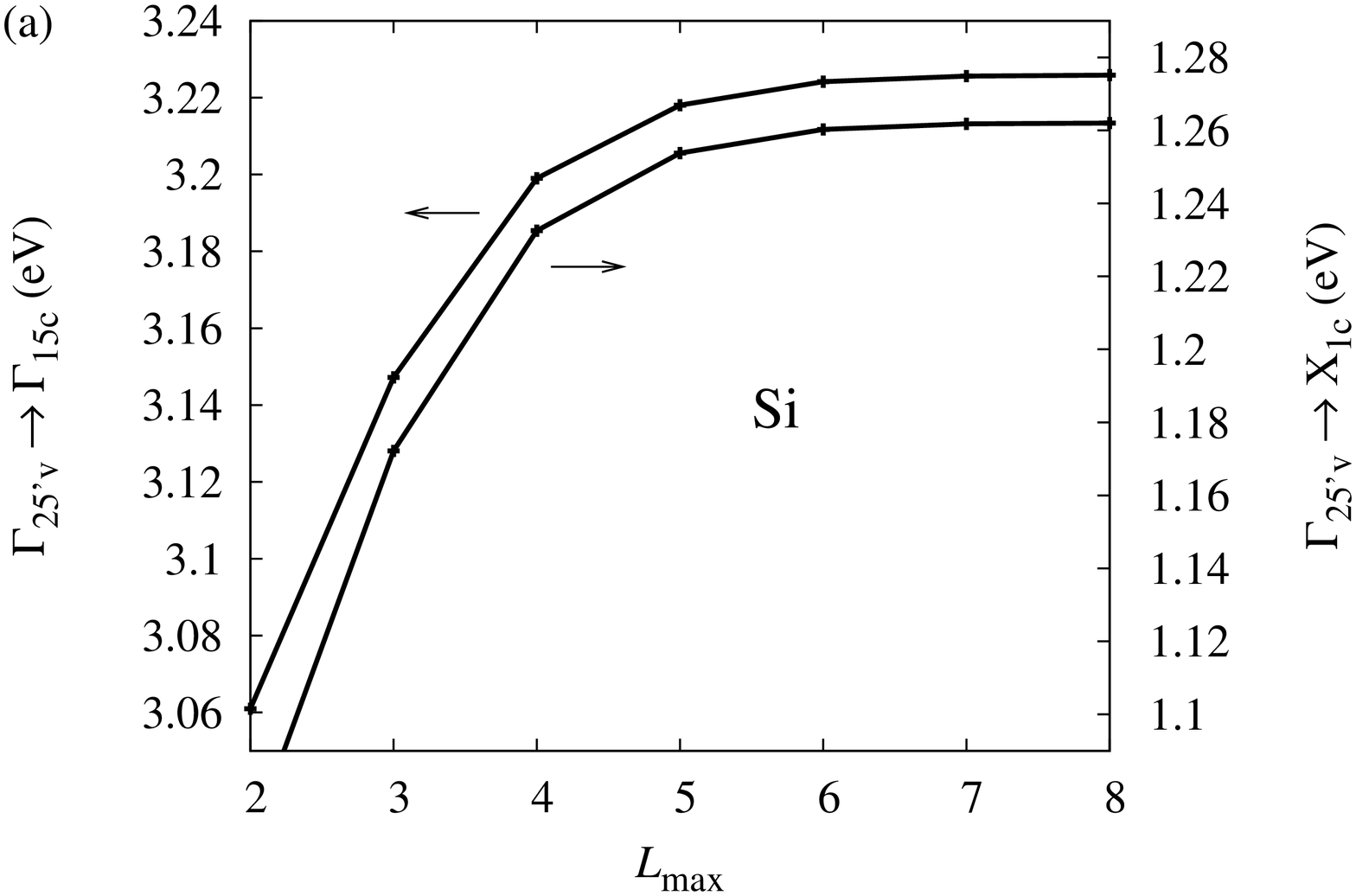}

\includegraphics[%
  width=0.70\columnwidth,
  angle=-90]{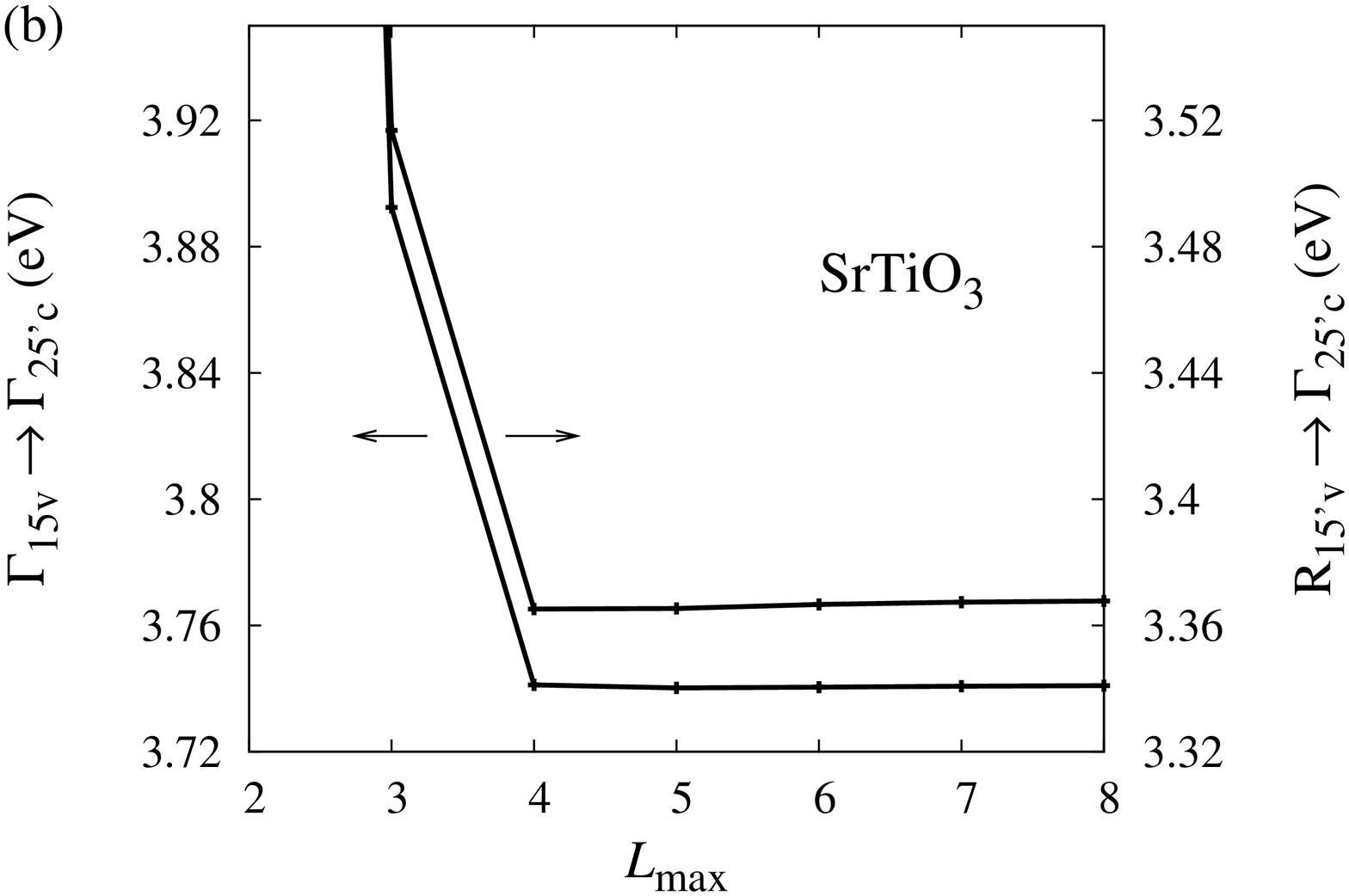}

\caption{\label{cap:Lcut-conv}Convergence of (a) the $\Gamma_{25'\mathrm{v}}\rightarrow\Gamma_{15\mathrm{c}}$
and $\Gamma_{25'\mathrm{v}}\rightarrow\mathrm{X}_{1\mathrm{c}}$ gaps
of Si and (b) the $\Gamma_{15\mathrm{v}}\rightarrow\Gamma_{25'\mathrm{c}}$
and $\mathrm{R}_{15'\mathrm{v}}\rightarrow\Gamma_{25'\mathrm{c}}$
gaps of $\mathrm{SrTiO}_{3}$ as a function of the MPB cutoff parameter
$L_{\mathrm{max}}$ for the angular quantum number. }
\end{figure}
The MPB is designed as a basis for the products of the wave functions
{[}Eq.~(\ref{Eq:FLAPW-basis}){]}, for which MT functions with angular
momenta as large as $l_{\mathrm{max}}=8$ or even larger are typically
taken into account. As a consequence, an exact representation of the
products inside the spheres requires spherical harmonics at least
up to $2l_{\mathrm{max}}$. However, the high angular momenta in the
original FLAPW basis are mostly needed to ensure an accurate matching
to the IPWs and contribute little to the actual wave functions. In
fact, we find that the cutoff parameter $L_{\mathrm{max}}$ for the
MPB can be chosen much smaller than $2l_{\mathrm{max}}$ and, indeed,
even smaller than $l_{\mathrm{max}}$. Figure \ref{cap:Lcut-conv}
shows the convergence of the quasiparticle transitions $\Gamma_{25'\mathrm{v}}\rightarrow\Gamma_{15\mathrm{c}}$
and $\Gamma_{25'\mathrm{v}}\rightarrow\mathrm{X}_{1\mathrm{c}}$ in
(a) $\mathrm{Si}$ as well as $\Gamma_{15\mathrm{v}}\rightarrow\Gamma_{25'\mathrm{c}}$
and $\mathrm{R_{15'v}}\rightarrow\Gamma_{25'\mathrm{c}}$ in (b) $\mathrm{SrTiO}_{3}$
with respect to $L_{\mathrm{max}}$. Convergence to within 0.01~eV
is already attained with $L_{\mathrm{max}}=5$ for Si and $L_{\mathrm{max}}=4$
for $\mathrm{SrTiO}_{3}$. With these cutoff values, the MPB contains
292 and 853 MT functions, respectively.

\begin{figure}
\includegraphics[%
  width=0.70\columnwidth,
  angle=-90]{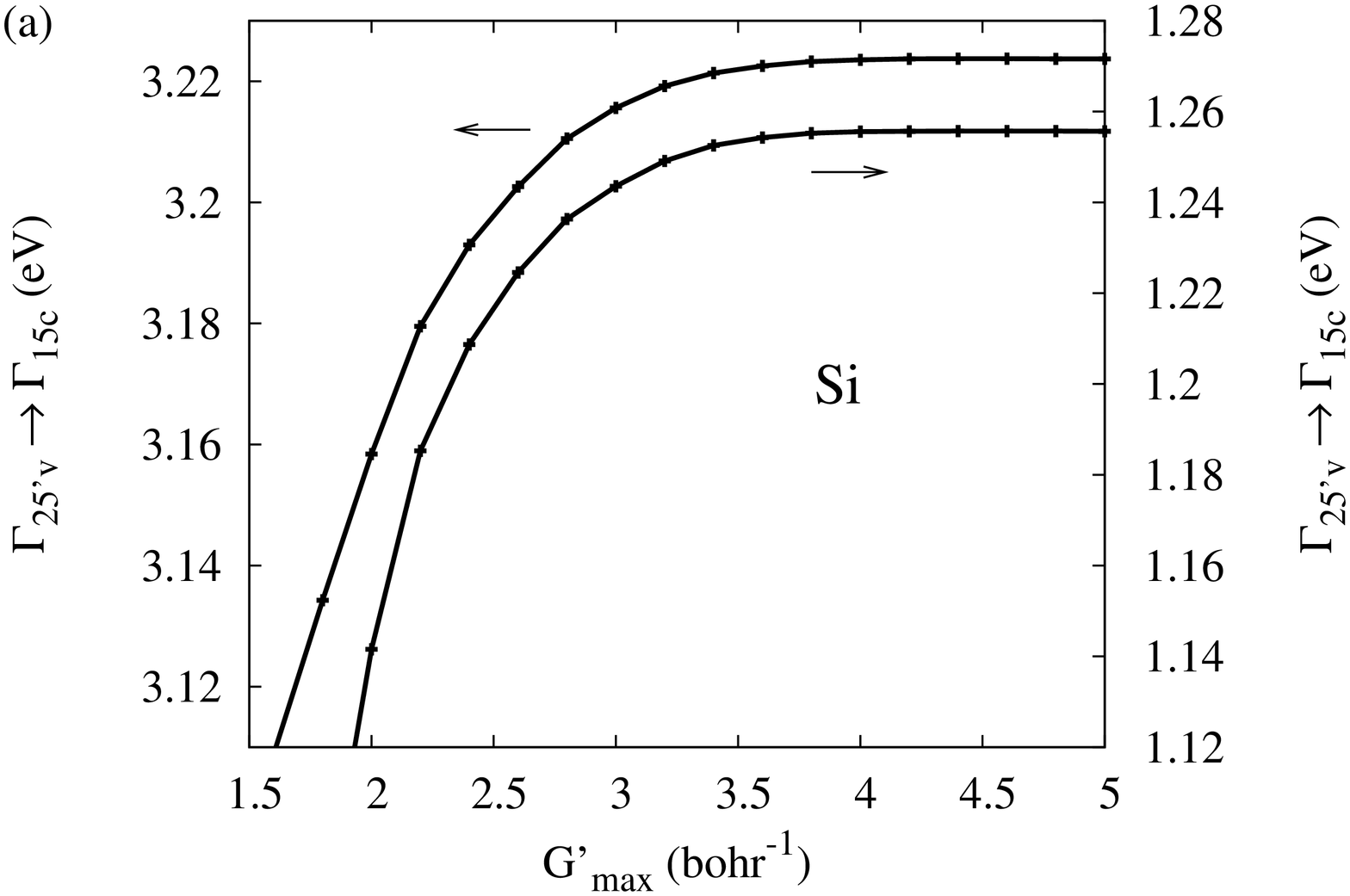}

\includegraphics[%
  width=0.70\columnwidth,
  angle=-90]{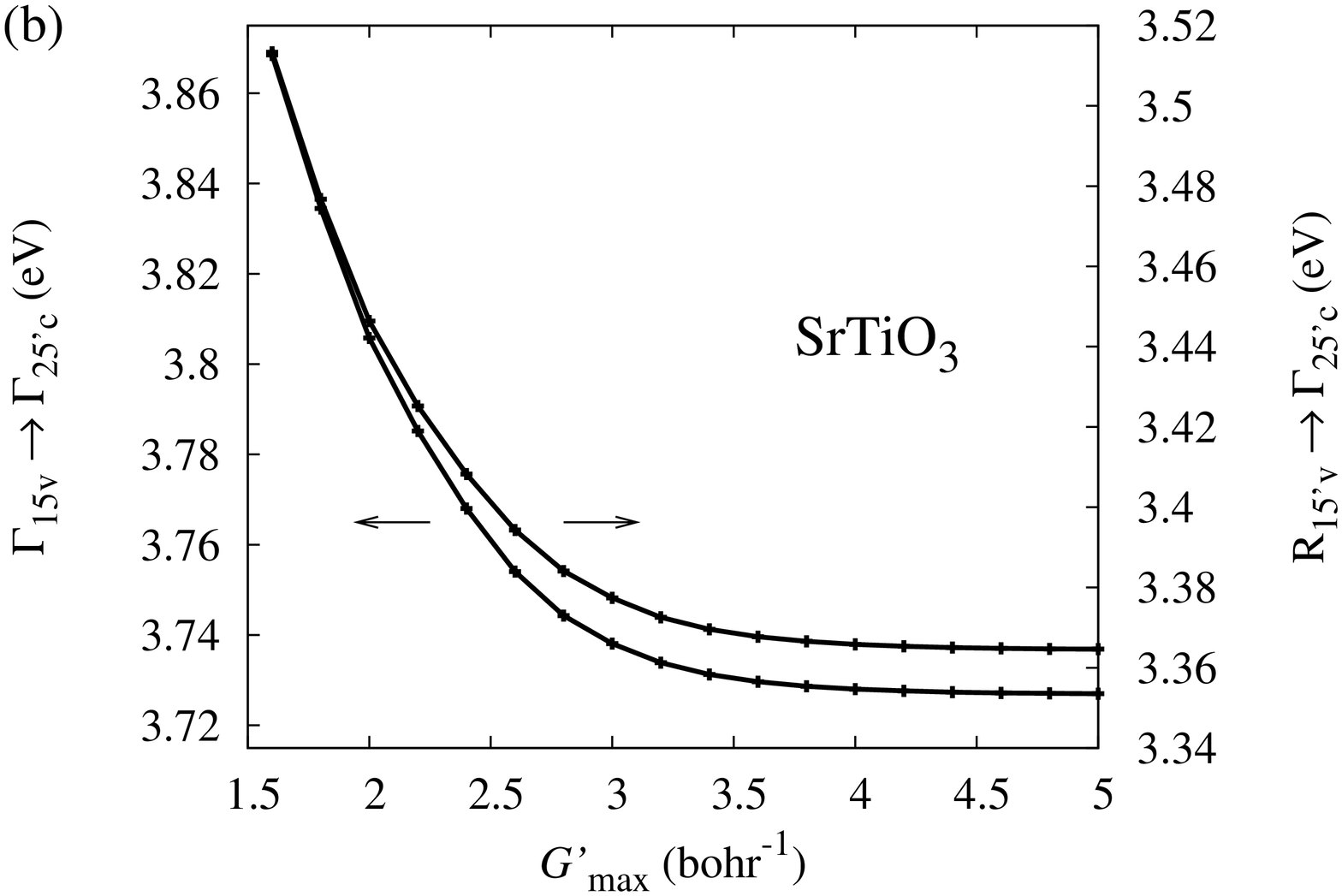}

\caption{\label{cap:Gcut-conv}Convergence of (a) the $\Gamma_{25'\mathrm{v}}\rightarrow\Gamma_{15\mathrm{c}}$
and $\Gamma_{25'\mathrm{v}}\rightarrow\mathrm{X}_{1\mathrm{c}}$ gaps
of Si and (b) the $\Gamma_{15\mathrm{v}}\rightarrow\Gamma_{25'\mathrm{c}}$
and $\mathrm{R_{15'\mathrm{v}}\rightarrow\Gamma_{25'\mathrm{c}}}$
gaps of $\mathrm{SrTiO}_{3}$ as a function of the MPB cutoff parameter
$G_{\mathrm{max}}^{\prime}$ for the IPWs. }
\end{figure}
A similar statement can be made about the convergence parameter $G_{\mathrm{max}}^{\prime}$
for the IPWs. Although an exact representation requires twice the
wave-function cutoff $G_{\mathrm{max}}$, which we choose as $3.6\,\mathrm{bohr}^{-1}$
for Si and $4.3\,\mathrm{bohr}^{-1}$ for $\mathrm{SrTiO}_{3}$ (giving
rise to around 200 and 550 augmented plane waves, respectively), again
a much smaller cutoff parameter is sufficient for the products: Figure
\ref{cap:Gcut-conv} confirms that convergence to within 0.01~eV
is achieved with $G'_{\mathrm{max}}=2.7\,\mathrm{bohr}^{-1}$ and
$G'_{\mathrm{max}}=3.2\,\mathrm{bohr}^{-1}$ for Si and $\mathrm{SrTiO}_{3}$,
respectively, which yield around 100 and 250 IPWs in the MPB. In general,
we find that the ratio $G_{\mathrm{max}}^{\prime}/G_{\mathrm{max}}\approx0.75$
can be used as a rule of thumb and works well for all materials considered
here.

\begin{figure}
\includegraphics[%
  width=0.70\columnwidth,
  angle=-90]{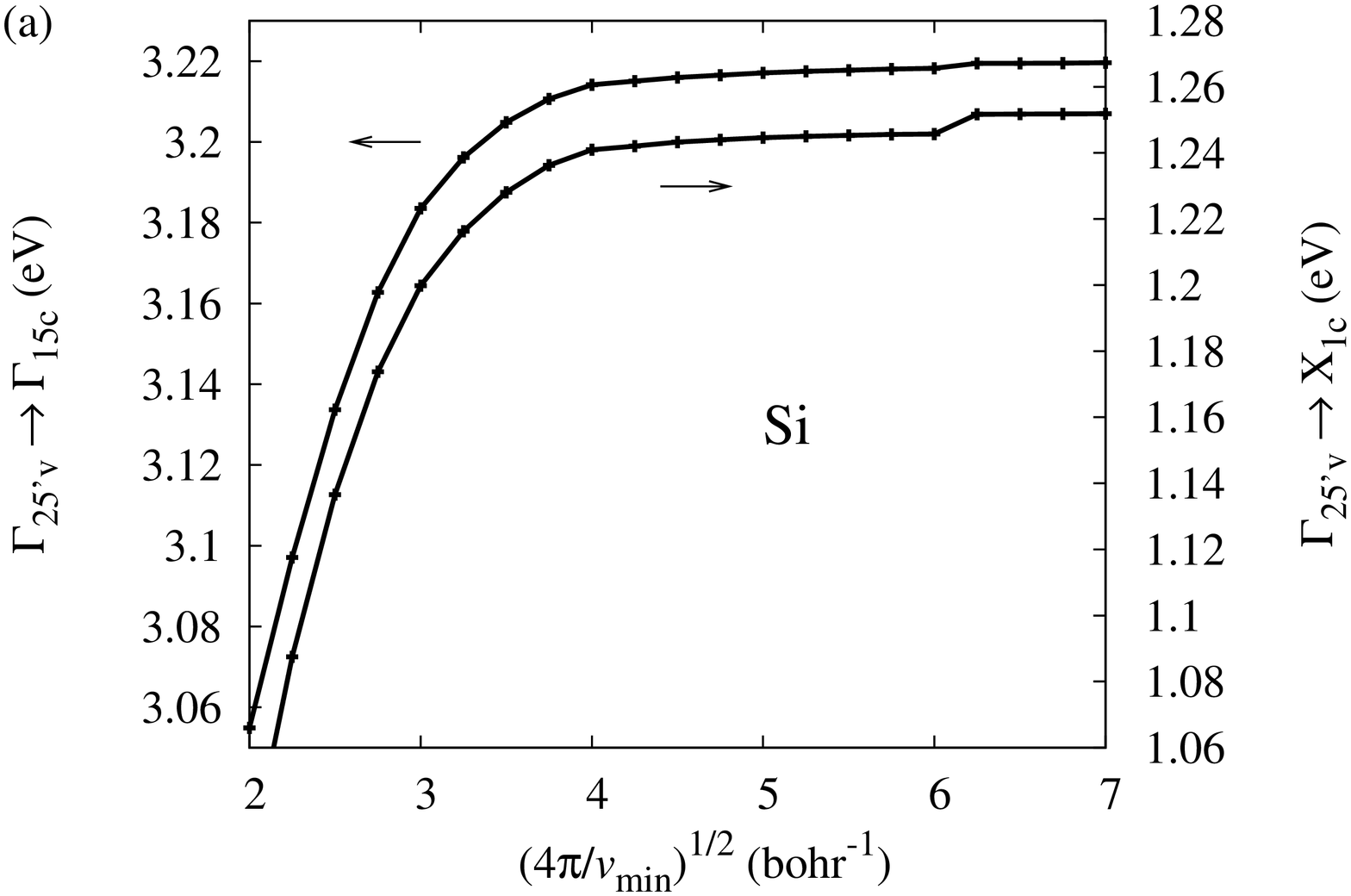}

\includegraphics[%
  width=0.70\columnwidth,
  angle=-90]{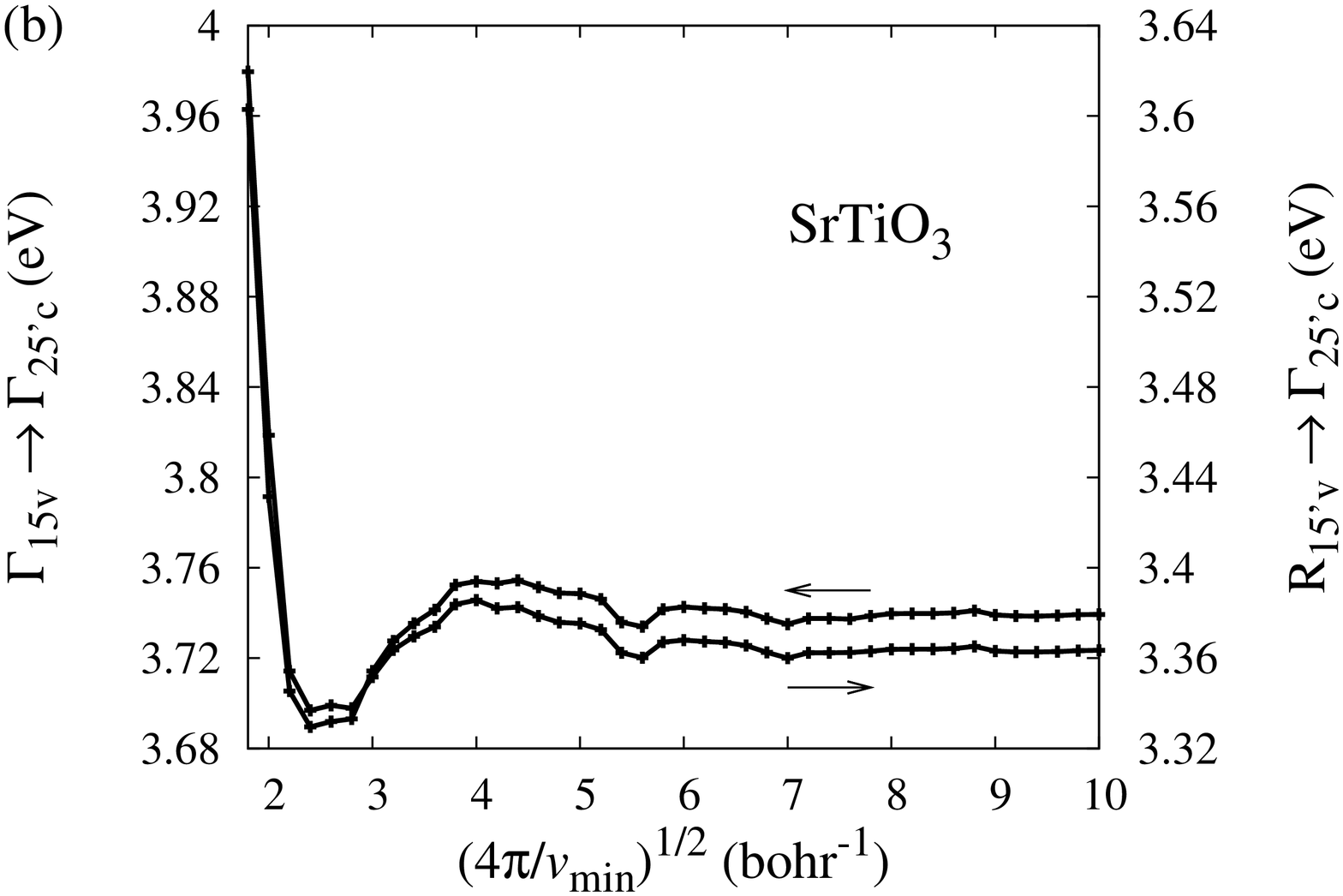}

\caption{\label{cap:mb-conv}Convergence of (a) the $\Gamma_{25'\mathrm{v}}\rightarrow\Gamma_{15\mathrm{c}}$
and $\Gamma_{25'\mathrm{v}}\rightarrow\mathrm{X}_{1\mathrm{c}}$ gaps
of Si and (b) the $\Gamma_{15\mathrm{v}}\rightarrow\Gamma_{25'\mathrm{c}}$
and $\mathrm{R_{\mathrm{v}}\rightarrow\Gamma_{25'\mathrm{c}}}$ gaps
of $\mathrm{SrTiO}_{3}$ as a function of $\sqrt{4\pi/v_{\mathrm{min}}}$. }
\end{figure}
\begin{figure}
\includegraphics[%
  width=0.70\columnwidth,
  angle=-90]{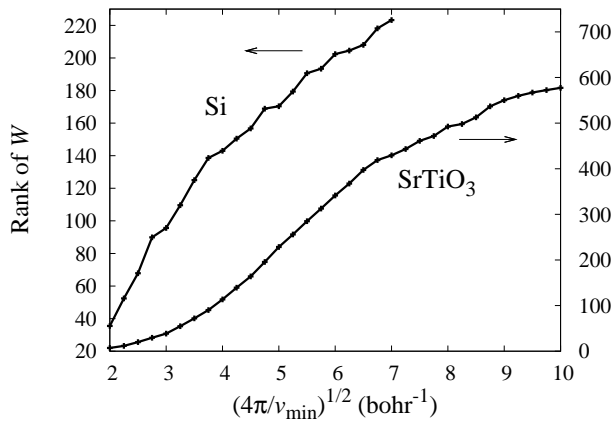}

\caption{\label{cap:mb-order}Rank of the matrix $W$ for the screened interaction,
which equals the number of basis functions after the optimization,
for Si and $\mathrm{SrTiO}_{3}$ as a function of $\sqrt{4\pi/v_{\mathrm{min}}}$. }
\end{figure}
In Fig.~\ref{cap:mb-conv}, we show the convergence of the gap energies
of Si and $\mathrm{SrTiO}_{3}$ with respect to the threshold value
$v_{\mathrm{min}}$ defined in Sec.~\ref{sub:OptimizationMPB}. If
the MPB was complete, the eigenvalues of the Coulomb matrix would
be given by the Fourier transform $v_{\mathbf{G}}(\mathbf{k})=4\pi/|\mathbf{k+G}|^{2}$.
With this in mind, the threshold value can be reformulated in terms
of a cutoff in reciprocal space $|\mathbf{k+G}|\le\sqrt{4\pi/v_{\mathrm{min}}}$,
very similar to that for the IPWs discussed above. Therefore, it is
reasonable to show the convergence in terms of this cutoff value,
even though the MPB is only complete in the subspace spanned by the
wave-function products, of course, and the Fourier transform $v_{\mathbf{G}}(\mathbf{k})$
is hence only an estimate for the eigenvalues. 

Figure \ref{cap:mb-conv}(a) shows the convergence of the quasiparticle
transitions for Si with respect to $\sqrt{4\pi/v_{\mathrm{min}}}$.
We observe that the values are converged to within 0.01~eV around
3.5~$\mathrm{bohr}^{-1}$, which corresponds to $v_{\mathrm{min}}=1.9\,\mathrm{ha}$.
With these values the rank of the matrix $W$ (see Fig.~\ref{cap:mb-order})
is reduced from 392 (the full MPB) to around 75 and the computation
time from 140 to 42s on an Intel Xeon (2.66 GHz, 4 MB cache) work
station. Interestingly, the curve of the indirect transition in Fig.~\ref{cap:mb-conv}(a)
exhibits a sudden step between 5.25 and 5.5~$\mathrm{bohr}^{-1}$,
where the gap energy changes by 4~meV. A similar but much smaller
step of 0.7~meV can also be observed in the direct transition. Noting
the simplified estimates $v_{\mathbf{G}}(\mathbf{k})$ for the eigenvalues,
this can be attributed to a shell of reciprocal vectors with length
$|\mathbf{k+G}|$ that enter between the radii 5.25 and 5.5~$\mathrm{bohr}^{-1}$
and give a sizable contribution. Although simplified, this is the
correct picture, because the true set of eigenvalues $v_{\mu}(\mathbf{k})$
usually contains many groups of degenerate eigenvalues, especially
at high-symmetry points $\mathbf{k}$ in the BZ. By analogy, these
groups can be viewed as shells of $\mathbf{k+G}$ vectors in reciprocal
space. 

In $\mathrm{SrTiO}_{3}$, the gap energies converge somewhat less
smoothly, but systematically. From Fig.~\ref{cap:mb-conv}(b), we
see that after 5.3~$\mathrm{bohr}^{-1}$ the energies change by less
than 0.01~eV. We note that the convergence criterion of 0.01~eV
is quite ambitious for $GW$ calculations. If we relax this criterion
to, e.g., 0.05~eV, which should be sufficient for most studies, considerably
smaller cutoffs suffice.

\begin{figure}
\includegraphics[%
  width=0.70\columnwidth,
  angle=-90]{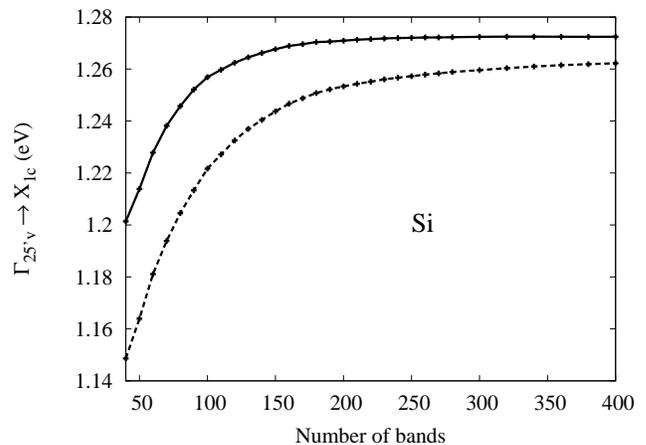}

\caption{\label{cap:nband}Convergence of the $\Gamma_{25'\mathrm{v}}\rightarrow\mathrm{X}_{1\mathrm{c}}$
gap of Si with respect to the number of bands used in the construction
of the polarization function as well as the self-energy with (solid
line) and without the extrapolar correction (dashed line). }
\end{figure}
Equations (\ref{eq:polar_MB}) and (\ref{eq:SEc_AC}) involve a summation
over the unoccupied states $\varphi_{n'\mathbf{q+k}}^{\sigma}(\mathbf{r)}$.
In practice, we must truncate this sum at some maximal band index
$N'$. It is well known that a proper convergence of the $GW$ quasiparticle
energies requires very many unoccupied states.\cite{Tiago2004,Friedrich2006}
While for $\mathrm{SrTiO}_{3}$, a relatively modest number of 200
states is sufficient, the band gaps of Si, in particular, the indirect
one, are more difficult to converge. Recently, Bruneval and Gonze\cite{Bruneval2008}
proposed an approximate scheme that corrects for the neglect of the
states with indices $n'>N'$ and only involves the states with $n'\le N'$.
They showed that this \emph{extrapolar} correction reduces the number
of states needed for convergence considerably with only a small computational
overhead. In short, all states $n'>N'$ are placed on a fixed energy
above all others, which allows to take the energy denominator of Eqs.~(\ref{eq:polar_MB})
and (\ref{eq:SEc_AC}) out of the sum over $n'$ and to use the completeness
relation for the one-particle states $\varphi_{n'\mathbf{q+k}}^{\sigma}(\mathbf{r)}$.
The final expression can then be evaluated only with the states $n'\le N'$.
We have implemented this scheme in an all-electron code. Contrary
to Ref.~\onlinecite{Bruneval2008} we do not employ a plasmon-pole
model, though, but use the full matrix of the screened interaction
in the correction. As shown in Fig.~\ref{cap:nband}, we find a considerably
improved convergence with respect to the number of bands, too. The
fixed energy for the bands $n'>N'$ is placed 16~Ry (217.7~eV) above
the maximal energy of the bands $n'\le N'$. However, all other results
in this paper were obtained with the conventional summation. We note
that the LAPW basis is a relatively small and accurate basis for the
occupied states. In order to get enough unoccupied states for $GW$
calculations it is therefore often necessary to extend the LAPW basis
by increasing the reciprocal cutoff radius $G_{\mathrm{max}}$ and
introducing additional local orbitals.

\begin{table}

\caption{\label{cap:Materials}Fundamental $GW$ band gaps for a variety of
semiconductors and insulators compared with experimental and theoretical
values from the literature. We also indicate the LDA eigenvalue gaps.
All values are in electron volts.}

\begin{ruledtabular}

\begin{tabular}{lrrrrrrl}
&
LDA&
$GW$&
LDA$^{\mathrm{a}}$&
$GW$$^{\mathrm{a}}$&
LDA$^{\mathrm{b}}$&
$GW$$^{\mathrm{b}}$&
Expt.\tabularnewline
\hline
Ge&
0.02&
0.75&
--0.08&
0.57&
---&
---&
0.74$^{\mathrm{c}}$\tabularnewline
Si&
0.62&
1.11&
0.46&
0.90&
0.62&
1.12&
1.17$^{\mathrm{d}}$\tabularnewline
GaAs&
0.29&
1.31&
0.33&
1.31&
0.49&
1.30&
1.63$^{\mathrm{d}}$\tabularnewline
CdS&
1.17&
2.18&
---&
---&
1.14&
2.06&
2.58$^{\mathrm{e}}$\tabularnewline
GaN&
1.67&
2.83&
1.81&
3.03&
1.62&
2.80&
3.27$^{\mathrm{f}}$\tabularnewline
$\mathrm{SrTiO}_{3}$&
1.80&
3.36&
---&
---&
---&
---&
3.25$^{\mathrm{g}}$\tabularnewline
$\mathrm{BaTiO}_{3}$&
2.18&
3.18&
---&
---&
---&
---&
3.3$^{\mathrm{h}}$\tabularnewline
CaSe&
2.04&
3.63&
---&
---&
---&
---&
3.85$^{\mathrm{i}}$\tabularnewline
C&
4.15&
5.62&
4.11&
5.49&
4.12&
5.50&
5.48$^{\mathrm{d}}$\tabularnewline
BN&
4.35&
6.20&
---&
---&
4.45&
6.10&
5.97$^{\mathrm{j}}$\tabularnewline
MgO&
4.64&
7.17&
4.85&
6.77&
4.76&
7.25&
7.83$^{\mathrm{k}}$\tabularnewline
NaCl&
4.90&
7.53&
---&
---&
---&
---&
8.5$^{\mathrm{l}}$\tabularnewline
\end{tabular}

\end{ruledtabular} 

\begin{flushleft}$^{\mathrm{a}}$Reference \onlinecite{Kotani2002}\\
$^{\mathrm{b}}$Reference \onlinecite{Shishkin2007}\\
$^{\mathrm{c}}$Reference \onlinecite{Macfarlane1957}\\
$^{\mathrm{d}}$Reference \onlinecite{LandBoern}\\
$^{\mathrm{e}}$Reference \onlinecite{Magnusson1987}\\
$^{\mathrm{f}}$Reference \onlinecite{Okumura1994}\\
$^{\mathrm{g}}$Reference \onlinecite{vanBenthem2001}\\
$^{\mathrm{h}}$Reference \onlinecite{Hudson1993}\\
$^{\mathrm{i}}$Reference \onlinecite{Kaneko1988}\\
$^{\mathrm{j}}$Reference \onlinecite{Watanabe2004}\\
$^{\mathrm{k}}$Reference \onlinecite{Whited1973}\\
$^{\mathrm{l}}$Reference \onlinecite{Poole1975}\end{flushleft}
\end{table}
For reference, we list the fundamental LDA and $GW$ band gaps for
a variety of semiconductors and insulators in Table \ref{cap:Materials},
together with experimental and other theoretical values for comparison.
The latter are calculated with the LMTO (Ref.~\onlinecite{Kotani2002})
and the PAW method (Ref.~\onlinecite{Shishkin2007}). Our own results
for the fundamental band gaps are converged to within 0.01~eV with
respect to the numerical parameters, including the BZ sampling. We
find that an accurate description of high-lying unoccupied states
with additional local orbitals is crucial for properly converged $GW$
results. The core states can also have a sizable effect on electron
correlation and the resulting band gaps. For example, inclusion of
the cation $2s$ and $2p$ states of MgO and NaCl changes their band
gaps by as much as 0.2~eV. Semicore states (e.g., Mg $2p$) are described
with local orbitals, while deeper core states (e.g., Mg $2s$) are
treated as dispersionless bands, whose wave functions are confined
to the MT spheres. Overall our LDA and $GW$ values agree very well
with those of Ref.~\onlinecite{Shishkin2007}, but somewhat less
so with the older Ref.~\onlinecite{Kotani2002}. As expected, the
LDA considerably underestimates the band gaps. The $GW$ self-energy
corrects this underestimation in such a way that the results come
very close to the measured values. However, there is still a slight
underestimation in most cases. It has been suggested that a self-consistent
scheme could improve the $GW$ values further.\cite{Shishkin2007,Kotani2007}
The starting point is then optimized in such a way that the resulting
one-particle orbitals are as close as possible to the quasiparticle
wave functions; in particular, closer than those from standard local
or semilocal functionals. In this way, the first-order perturbative
correction {[}Eq.~(\ref{eq:quasipart_corr}){]}, where the quasiparticle
wave functions are approximated by the one-particle orbitals, is better
justified. However, self-consistent $GW$ calculations are computationally
very expensive. When compared with the electronic self-energy, the
most obvious source of errors in the local and semilocal DFT functionals
is the missing self-interaction correction, which influences the shape
of the KS wave functions. Therefore, better results might alternatively
be obtained if one uses a functional that treats electronic exchange
more accurately, e.g., the exact exchange functional within the optimized-effective-potential
method or hybrid functionals.\cite{Rinke2005,Bechstedt2009} These
approaches go beyond the scope of the present paper. Nevertheless,
we note that the numerical procedure for the $GW$ approximation presented
here is independent of the starting point and could also be applied
within a self-consistent scheme or to functionals containing an exact
exchange term.

As the $GW$ approximation contains the exact exchange self-energy,
it does not suffer from the unphysical self-interaction error present
in local density functionals such as the LDA. Localized states are
most strongly affected by this error and appear too high in energy
within the LDA. Thus, the absence of the self-interaction error in
the $GW$ approximation should lead to an improved description of
these states. In fact, the quasiparticle levels of the localized Ga
and As $3d$ semicore states in gallium arsenide lie 2.1 and 3.1~eV
deeper than their LDA counterparts. Their theoretical binding energies
are 16.9 and 38.4 eV, respectively, which still underestimates the
experimental values of 18.82 and 40.76~eV from x-ray photoemission
spectroscopy.\cite{Ley1974} It has been shown that self-consistent
calculations can further improve the $d$-band positions.\cite{Kotani2007,Shishkin2007}

\begin{figure}
\includegraphics[%
  scale=0.46,
  angle=-90]{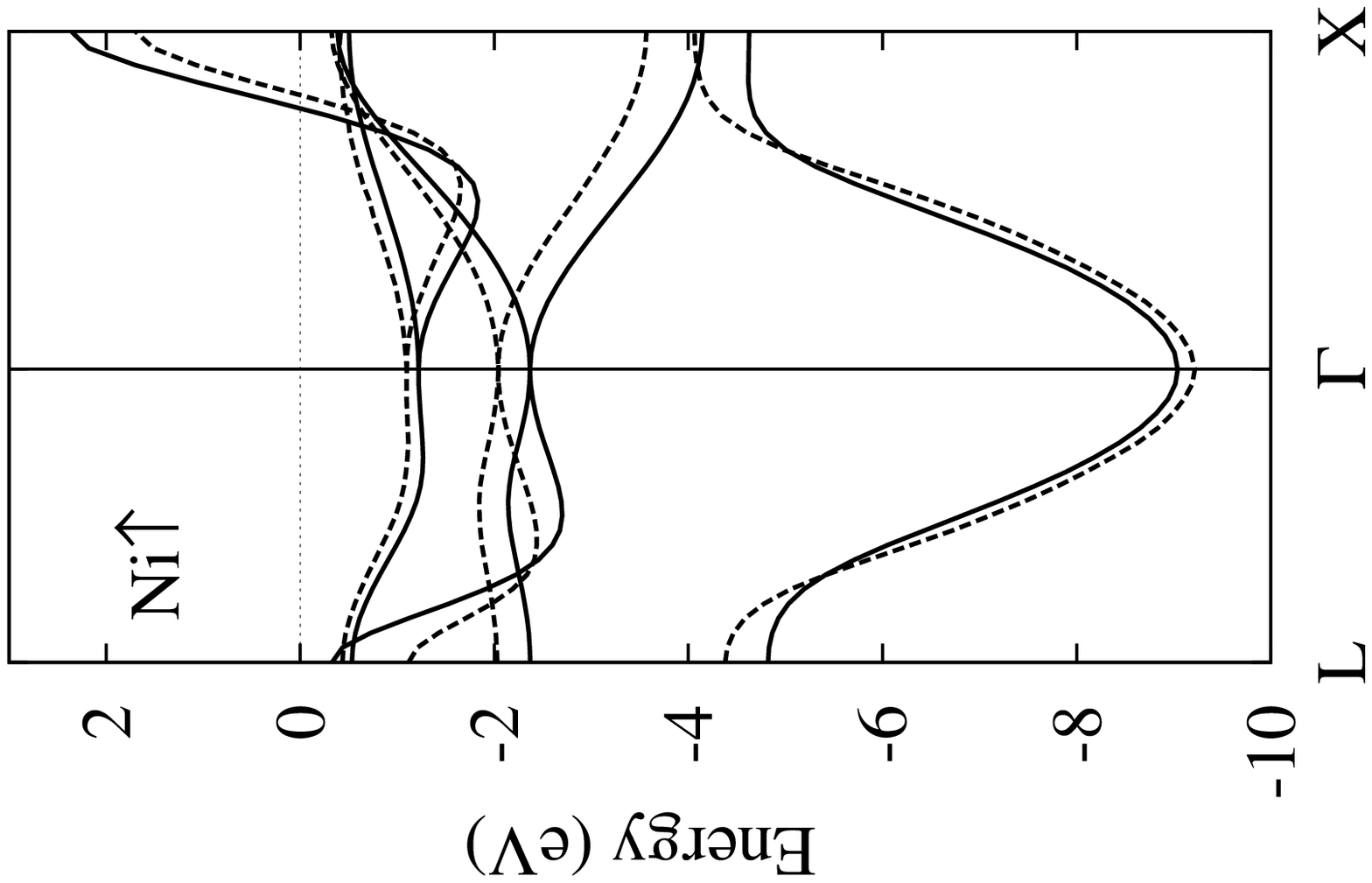} \includegraphics[%
  scale=0.46,
  angle=-90]{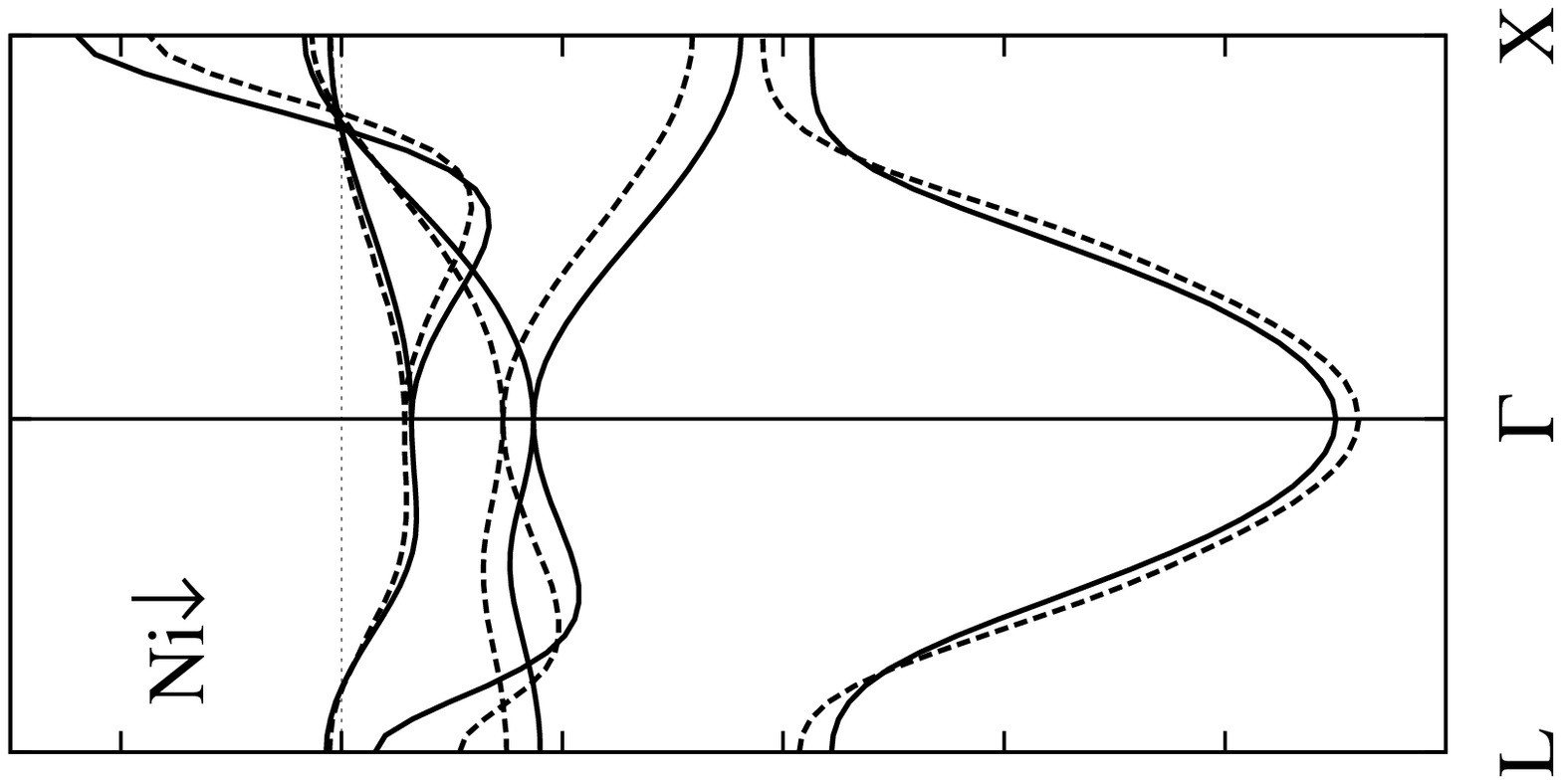}

\caption{\label{cap:Ni}Band structure of Ni calculated by the LSDA (dashed
lines) and the $GW$ approximation (solid lines) for majority (left
panel) and minority spin (right panel). The Fermi energy is set to
zero.}
\end{figure}
In Fig.~\ref{cap:Ni}, we show the local-spin-density approximation
(LSDA) (Ref.~\onlinecite{Perdew1981}) and $GW$ band structures
for ferromagnetic Ni. The self-energy correction was calculated with
a 8$\times$8$\times$8 sampling of the BZ. Convergence was tested
with a 10$\times$10$\times$10 set. While for the semiconductors
and insulators treated so far a model function {[}Eq.~(\ref{eq:modelf}){]}
for the correlation self-energy with three poles, i.e., $N_{\mathrm{p}}=3$,
was sufficient, we must use a five-pole function in the case of Ni
to reproduce the values that we get from the reference contour-integration
method. A comparison of the LSDA and $GW$ band structures shows that
the self-energy correction is strongly state and $\mathbf{k}$ dependent
whereas in the case of materials with a band gap, the quasiparticle
shifts are more or less uniform over the BZ but different for occupied
and unoccupied states. The $GW$ quasiparticle correction reduces
the occupied $d$-band width from about 4.0~eV in the LSDA to 3.2~eV,
which is in accordance with x-ray photoemission experiments.\cite{Hoechst1977}
On the other hand, the exchange splitting is hardly improved. There
is only a slight reduction, which cannot account for the large overestimation
within LSDA. The reason for this shortcoming is that the $GW$ self-energy
lacks two-particle vertex corrections, which give rise to spin-dependent
screening and thus to a different correction for spin-up and spin-down
states. Furthermore, the 6~eV satellite, which originates from a
virtual bound two-hole excitation, cannot be described within the
$GW$ approximation for the same reason. Our $GW$ band structure
compares very well with an early work within the LAPW method\cite{Aryasetiawan1992}
but less well with a more recent LMTO calculation.\cite{Yamasaki2003}
About this discrepancy we can only speculate. It might be attributed
to a less accurate description of unoccupied states within the LMTO
basis or to the usage of the offset-$\Gamma$ method in Ref.~\onlinecite{Yamasaki2003},
in which the numerically important region around the center of the
BZ is treated only approximately.

\begin{table}

\caption{\label{cap:Diamond}Computational time for the calculation of quasiparticle
shifts of diamond in the conventional (1$\times$1$\times$1) and
supercell (2$\times$2$\times$2 and 4$\times$4$\times$4) geometries.
For the former we also show corresponding timings without use of symmetry
({}``No'') and with restricted use ({}``Only IS'' and {}``Only
IBZ'') as well as the effect of the MPB optimization with a threshold
value $v_{\mathrm{min}}$. }

\begin{ruledtabular}

\begin{tabular}{rrrrrr}
Geometry&
Atoms&
\textbf{k} mesh&
Symmetry&
$v_{\mathrm{min}}$&
CPU time\tabularnewline
\hline
1$\times$1$\times$1&
2&
4$\times$4$\times$4&
No&
---&
5 min 52 s\tabularnewline
&
&
&
Only IS&
---&
3 min 34 s\tabularnewline
&
&
&
Only IBZ&
---&
30 s\tabularnewline
&
&
&
Yes&
---&
11 s\tabularnewline
&
&
&
Yes&
0.65&
5 s\tabularnewline
2$\times$2$\times$2&
16&
2$\times$2$\times$2&
Yes&
0.65&
14 min 15 s\tabularnewline
4$\times$4$\times$4&
128&
1$\times$1$\times$1&
Yes&
0.65&
34 h 11 min\tabularnewline
\end{tabular}

\end{ruledtabular} 
\end{table}
In order to demonstrate the efficiency of the code, we show the computational
time for calculating quasiparticle shifts for diamond in the conventional
unit cell (1$\times$1$\times$1) containing two atoms as well as
in 2$\times$2$\times$2 and 4$\times$4$\times$4 supercell geometries
containing 16 and 128 atoms, respectively. We choose the parameters
so that all three calculations yield identical results. For example,
the $\mathbf{k}$ mesh contains 4$\times$4$\times$4, 2$\times$2$\times$2,
and 1$\times$1$\times$1 points, respectively. The other parameters
are determined to ensure convergence to within 0.01~eV. The computation
times on a single CPU are given in Table \ref{cap:Diamond}. While
the calculation of the quasiparticle shifts takes only 5~s for the
conventional unit cell, even the treatment of supercells containing
16 and 128 atoms only consumes affordable 0.24 and 34.2~h of computation
time, respectively. 

In the case of the conventional unit cell (1$\times$1$\times$1),
we also demonstrate the efficiency gain achieved by exploiting the
symmetry according to Sec.~\ref{sub:Symmetry} and by using a threshold
parameter $v_{\mathrm{min}}$ as in Sec.~\ref{sub:OptimizationMPB}.
If symmetry is not used at all, then the computation of the quasiparticle
shifts takes nearly 6~min. The diamond structure exhibits inversion
symmetry (IS), which allows to define the bare and screened Coulomb
matrices as real symmetric instead of complex Hermitian quantities
after a symmetrization of the MT functions as briefly described in
Sec.~\ref{sec:mixedbasis} (for more details, see Ref.~\onlinecite{Betzinger}).
This reduces the computation time roughly by a factor of 2 ({}``Only
IS''). If we next calculate the screened interaction only in the
irreducible wedge of the BZ, i.e., at eight $\mathbf{k}$ points instead
of 64, the computation time goes down further to 30~s ({}``Only
IBZ''). The gain is slightly less than a factor of 8 because the
BZ summation in Eq.~(\ref{eq:SEc_MB}) must still be performed with
all 64 $\mathbf{k}$ points. We can only restrict this summation and
the sum over $\mathbf{q}$ in Eq.~(\ref{eq:polar_MB}) if we use
the extended IBZ (EIBZ) as explained in Sec.~\ref{sub:Symmetry},
which leads to further time savings of a factor of 3. Compared to
the first calculation, the usage of symmetry thus leads to a 32 times
faster execution without loss of accuracy. By introducing a threshold
parameter $v_{\mathrm{min}}=0.65$ for the optimization of the MPB,
we can even reduce the computation time further to only 5~s, gaining
an overall factor of 70.

\section{Conclusions\label{sec:Conclusions}}

In this paper, we described an implementation of the $GW$ approximation
for the electronic self-energy within the all-electron full-potential
linearized augmented-plane-wave method.\cite{Fleur} We employ a mixed
product basis, which is specifically designed for the representation
of wave-function products and retains the full accuracy of the all-electron
framework. As all-electron $GW$ calculations have so far been prohibitive
for large systems due to the computational cost, we presented ways
to speed up the calculations considerably so that supercell calculations
for defect systems, nanowires, interface, or surface structures become
feasible. As a demonstration, we showed that our computer code can
treat 128 carbon atoms in a diamond supercell. This was achieved by
exploiting spatial and time-reversal symmetries in the evaluation
of the polarization function and the self-energy. Both quantities
exhibit a $\mathbf{k}$ dependence and also involve a BZ summation.
While we only need to consider $\mathbf{k}$ points in the IBZ for
the former, the latter can be restricted to an EIBZ, which accelerates
the code considerably. If the system exhibits inversion symmetry,
a symmetrization of the MT part of the mixed product basis leads to
real symmetric instead of complex Hermitian response matrices, which
reduces the CPU time and memory demand. Furthermore, for the correlation
part of the self-energy we can apply an optimization of the mixed
product basis that involves a basis transformation to the eigenvectors
of the Coulomb matrix. By neglecting eigenvectors with eigenvalues
below a certain threshold value we only retain the dominant part of
the bare electron-electron interaction. The threshold value then becomes
a convergence parameter. This optimization reduces the matrix sizes
of response quantities such as the screened interaction, again giving
rise to a speed up of the calculation. We note that no further approximations
such as plasmon-pole models or a range separation of the interaction
potential are introduced, and the anisotropy of the screening at $\mathbf{k=0}$
is fully taken into account. The divergence of the bare and the screened
interaction potential in the limit $\mathbf{k\rightarrow0}$ is treated
analytically while zeroth-order correction are derived with the help
of $\mathbf{k\cdot p}$ perturbation theory. This procedure gives
rise to a fast $\mathbf{k}$-point convergence, which is particularly
important for $GW$ calculations. 

We showed convergence tests for silicon and strontium titanate as
a prototype semiconductor and transition-metal oxide, respectively,
to illustrate the accuracy of the mixed product basis and its optimization
with a threshold value for the Coulomb eigenvalues. The results already
converge with relatively modest parameters. For example, for the angular
momenta inside the MT spheres and the plane-wave representation in
the IR, cutoff values well below the exact limit (i.e., twice the
corresponding FLAPW cutoffs) are sufficient for convergence of the
gap energies to within 0.01~eV. In fact, the cutoff values can even
be chosen smaller than the FLAPW cutoffs. For reference, we reported
the fundamental $GW$ band gaps for a variety of semiconductors and
insulators. Our results are in good agreement with recent $GW$ calculations
based on the PAW method and with experiments, although there is a
somewhat larger discrepancy with older $GW$ results obtained within
the LMTO method. For ferromagnetic Ni, we find that the $GW$ self-energy
reduces the $d$-band width from 4.0 to 3.2~eV in very good agreement
with experiment, but hardly improves the overestimation of the exchange
splitting within LSDA. These results are in accordance with previous
calculations. 

For simplicity, we have restricted ourselves to the non-self-consistent
approach. However, with the numerical procedure presented here we
are prepared to follow Ref.~\onlinecite{Kotani2007} and extend the
method to the quasiparticle self-consistent scheme. Within this approach,
the full self-energy matrix including off-diagonal elements is needed.
The extension of the numerical procedure developed in the present
paper to these elements is straightforward. 

\begin{acknowledgments}
The authors acknowledge valuable discussions with Markus Betzinger,
Andreas Gierlich, Gustav Bihlmayer, Takao Kotani, Mark van Schilfgaarde,
and Tatsuya Shishidou as well as financial support from the Deutsche
Forschungsgemeinschaft through the Priority Program 1145.
\end{acknowledgments}


\begin{thebibliography}{10}
\bibitem{Aulbur2000}W. G. Aulbur, L. Jönsson, and J. W. Wilkins, Solid State Phys. \textbf{54},
1 (1999).
\bibitem{Adler1962}S. L. Adler, Phys. Rev. \textbf{126}, 413 (1962); N. Wiser, \emph{ibid}.
\textbf{129}, 62 (1963).
\bibitem{Hedin1965}L. Hedin, Phys. Rev. \textbf{139}, A796 (1965).
\bibitem{Hybertsen1985}M. S. Hybertsen and S. G. Louie, Phys. Rev. Lett. \textbf{55}, 1418
(1985). 
\bibitem{Godby1986}R. W. Godby, M. Schl\"{u}ter, and L. J. Sham, Phys. Rev. Lett. \textbf{56},
2415 (1986). 
\bibitem{Hybertsen1986}M. S. Hybertsen and S. G. Louie, Phys. Rev. B \textbf{34}, 5390 (1986). 
\bibitem{Godby1987}R. W. Godby, M. Schl\"{u}ter, and L. J. Sham, Phys. Rev. B \textbf{35},
4170 (1987). 
\bibitem{Hamada1990}N. Hamada, M. Hwang, and A. J. Freeman, Phys. Rev. B \textbf{41},
3620 (1990).
\bibitem{Aryasetiawan1992}F. Aryasetiawan, Phys. Rev. B \textbf{46}, 13051 (1992).
\bibitem{Ku2002}W. Ku and A. G. Eguiluz, Phys. Rev. Lett. \textbf{89}, 126401 (2002). 
\bibitem{Usuda2002}M. Usuda, N. Hamada, T. Kotani, and M. van Schilfgaarde, Phys. Rev.
B \textbf{66}, 125101 (2002).
\bibitem{Kotani2002}T. Kotani and M. van Schilfgaarde, Solid State Commun. \textbf{121},
461 (2002).
\bibitem{Yamasaki2002}A. Yamasaki and T. Fujiwara, Phys. Rev. B \textbf{66}, 245108 (2002).
\bibitem{Faleev2004}S. V. Faleev, M. van Schilfgaarde, and T. Kotani, Phys. Rev. Lett.
\textbf{93}, 126406 (2004). 
\bibitem{Arnaud2000}B. Arnaud and M. Alouani, Phys. Rev. B \textbf{62}, 4464 (2000).
\bibitem{Lebegue2003}S. Leb\`{e}gue, B. Arnaud, M. Alouani, and P. E. Bloechl, Phys. Rev.
B \textbf{67}, 155208 (2003). 
\bibitem{Shishkin2006}M. Shishkin and G. Kresse, Phys. Rev. B \textbf{74}, 035101 (2006).
\bibitem{Ernst2005}A. Ernst, M. Lüders, P. Bruno, W. M. Temmerman, and Z. Szotek (unpublished).
\bibitem{Puschnig2002}P. Puschnig and C. Ambrosch-Draxl, Phys. Rev. B \textbf{66}, 165105
(2002).
\bibitem{Friedrich2009}C. Friedrich, A. Schindlmayr, and S. Blügel, Comput. Phys. Commun.
\textbf{180}, 347 (2009).
\bibitem{Kohn1965}W. Kohn and L. J. Sham, Phys. Rev. \textbf{140}, A1133 (1965).
\bibitem{Andersen1975}O. K. Andersen, Phys. Rev. B \textbf{12}, 3060 (1975).
\bibitem{Singh1991}D. Singh, Phys. Rev. B \textbf{43}, 6388 (1991).
\bibitem{Friedrich2006}C. Friedrich, A. Schindlmayr, S. Blügel, and T. Kotani, Phys. Rev.
B \textbf{74}, 045104 (2006).
\bibitem{Fleur}Further details available from http://www.flapw.de .
\bibitem{Aryasetiawan1994}F. Aryasetiawan and O. Gunnarsson, Phys. Rev. B \textbf{49}, 16214
(1994).
\bibitem{Betzinger}M. Betzinger, C. Friedrich, and S. Blügel (unpublished).
\bibitem{Dagens1972}L. Dagens and F. Perrot, Phys. Rev. B \textbf{5}, 641 (1972).
\bibitem{Rojas1995}H. N. Rojas, R. W. Godby, and R. J. Needs, Phys. Rev. Lett. \textbf{74},
1827 (1995); M. M. Rieger, L. Steinbeck, I. D. White, H. N. Rojas,
and R. W. Godby, Comput. Phys. Commun. \textbf{117}, 211 (1999).
\bibitem{Godby1988}R. W. Godby, M. Schlüter, and L. J. Sham, Phys. Rev. B \textbf{37},
10159 (1988); F. Aryasetiawan, in \emph{Electronic Structure Calculations}
in \emph{Advances in Condensed Matter Science}, edited by V. I. Anisimov
(Gordon and Breach, New York, 2000).
\bibitem{Rath1975}J. Rath and A. J. Freeman, Phys. Rev. B \textbf{11}, 2109 (1975).
\bibitem{Baroni1986}S. Baroni and R. Resta, Phys. Rev. B \textbf{33}, 7017 (1986).
\bibitem{Massidda1993}S. Massidda, M. Posternak, and A. Baldereschi, Phys. Rev. B \textbf{48},
5058 (1993).
\bibitem{Freysoldt2007}C. Freysoldt, P. Eggert, P. Rinke, A. Schindlmayr, R. W. Godby, and
M. Scheffler, Comput. Phys. Commun. \textbf{176}, 1 (2007).
\bibitem{Ziesche1983}\emph{Ergebnisse in der Elektronentheorie der Metalle}, edited by
P. Ziesche and G. Lehmann (Akademie/Springer, Berlin, 1983).
\bibitem{Kim}S. K. Kim, \emph{Group Theoretical Methods and Applications to Molecules
and Crystals} (Cambridge University Press, Cambridge, 1999).
\bibitem{Perdew1981}J. P. Perdew and A. Zunger, Phys. Rev. B \textbf{23}, 5048 (1981). 
\bibitem{Tiago2004}M. L. Tiago, S. Ismail-Beigi, and S. G. Louie, Phys. Rev. B \textbf{69},
125212 (2004).
\bibitem{Bruneval2008}F. Bruneval and X. Gonze, Phys. Rev. B \textbf{78}, 085125 (2008).
\bibitem{Shishkin2007}M. Shishkin and G. Kresse, Phys. Rev. B \textbf{75}, 235102 (2007).
\bibitem{Macfarlane1957}G. G. Macfarlane, T. P. McLean, J. E. Quarrington, and V. Roberts,
Phys. Rev. \textbf{108}, 1377 (1957).
\bibitem{LandBoern}T. C. Chiang and F. J. Himpsel, in \emph{Electronic Structure of Solids:
Photoemission Spectra and Related Data}, Landolt-Börnstein \emph{}New
Series, Group III \emph{}Vol. 23A, edited by A. Goldmann and E.-E.
Koch (Springer, Berlin, 1989).
\bibitem{Magnusson1987}K. O. Magnusson, U. O. Karlsson, D. Straub, S. A. Flodström, and F.
J. Himpsel, Phys. Rev. B \textbf{36}, 6566 (1987).
\bibitem{Okumura1994}H. Okumura, S. Yoshida, and T. Okahisa, Appl. Phys. Lett. \textbf{64},
2997 (1994).
\bibitem{vanBenthem2001}K. van Benthem, C. Elsässer, and R. H. French, J. Appl. Phys. \textbf{90},
6156 (2001).
\bibitem{Hudson1993}L. T. Hudson, R. L. Kurtz, S. W. Robey, D. Temple, and R. L. Stockbauer,
Phys. Rev. B \textbf{47}, 1174 (1993).
\bibitem{Kaneko1988}Y. Kaneko and T. Koda, J. Cryst. Growth \textbf{86}, 72 (1990).
\bibitem{Watanabe2004}K. Watanabe, T. Taniguchi, and H. Kanda, Nature Mater. \textbf{3},
404 (2004).
\bibitem{Whited1973}R. C. Whited, C. J. Flaten, and W. C. Walker, Solid State Commun.
\textbf{13}, 1903 (1973).
\bibitem{Poole1975}R. T. Poole, J. G. Jenkin, J. Liesegang, and R. C. G. Leckey, Phys.
Rev. B \textbf{11}, 5179 (1975).
\bibitem{Kotani2007}T. Kotani, M. van Schilfgaarde, and S. V. Faleev, Phys. Rev. B \textbf{76},
165106 (2007).
\bibitem{Rinke2005}P. Rinke, A. Qteish, J. Neugebauer, C. Freysoldt, and M. Scheffler,
New J. Phys. \textbf{7}, 126 (2005).
\bibitem{Bechstedt2009}F. Bechstedt, F. Fuchs, and G. Kresse, Phys. Status Solidi B \textbf{246},
1877 (2009).
\bibitem{Ley1974}L. Ley, R. A. Pollak, F. R. McFeely, S. P. Kowalczyk, and D. A. Shirley,
Phys. Rev. B \textbf{9}, 600 (1974).
\bibitem{Hoechst1977}H. Höchst, S. Hüfner, and A. Goldmann, Z. Phys. B \textbf{26}, 133
(1977).
\bibitem{Yamasaki2003}A. Yamasaki and T. Fujiwara, J. Phys. Soc. Jpn. \textbf{72}, 607 (2003).\end{thebibliography}
\end{document}